\documentclass[useAMS,usenatbib,referee]{mn2e}

\usepackage{graphicx,color}

\newcommand{\etal }{{et al.} }
\newcommand{\msun}{\thinspace M_\odot}

\newcommand{\rsun}{\thinspace R_\odot} 
\newcommand{\vect}[1]{\mbox{\boldmath$#1$}}
\def\lesssim{\mathrel{\hbox{\rlap{\hbox{\lower4pt\hbox{$\sim$}}}\hbox{$<$}}}}
\def\gtrsim{\mathrel{\hbox{\rlap{\hbox{\lower4pt\hbox{$\sim$}}}\hbox{$>$}}}}
\newcommand{\cm}{\,{\rm cm}^{-3} } 
\newcommand{\km}{\,{\rm km\, s}^{-1}}

\newcommand{\mdot}{M_\odot\,{\rm yr}^{-1} }
\newcommand{\tc}{t_{\rm c}}
\newcommand{\rhoc}{\rho_{\rm c}}
\newcommand{\dfrac}[2]{{\displaystyle \frac{#1}{#2}} }

\title[Formation of Population III Stars]{The Formation of Population III Stars in Gas Accretion Stage: Effects of Magnetic Fields}
\author[M. N. ~Machida \&   K. Doi]
  { Masahiro N. Machida$^{1}$\thanks{E-mail: machida.masahiro.018@m.kyushu-u.ac.jp (MNM); dn121001@center.konan-u.ac.jp (KD) } and  Kentaro Doi$^{2}$ \\
$^{1}$Department of Earth and Planetary Sciences, Faculty of Sciences, Kyushu University, 6-10-1 Hakozaki, Higashi-ku, Fukuoka, Japan\\
$^{2}$Department of Physics, Konan University, Okamoto 8-9-1, Kobe, Japan\\
}

\begin{document}
\maketitle
\begin{abstract}
The formation of Population III stars  is investigated using resistive magnetohydrodynamic simulations.
Starting from a magnetized primordial prestellar cloud, we calculate the cloud evolution several hundreds of years after first protostar formation, resolving the protostellar radius. 
When the natal minihalo field strength is weaker than $B \lesssim 10^{-13} (n/1\cm)^{-2/3}$\,G ($n$ is the hydrogen number density), magnetic effects can be ignored. 
In this case, fragmentation occurs frequently and a stellar cluster forms, in which stellar mergers and mass exchange between protostars  contribute to the mass growth of these protostars. 
During the early gas accretion phase, the most massive protostar remains near the cloud centre, whereas some of the less massive protostars are ejected.
The magnetic field significantly affects Population III star formation when $B_{\rm amb} \gtrsim 10^{-12} (n/1\cm)^{-2/3}$\,G.
In this case, because the angular momentum around the protostar is effectively transferred by both magnetic braking and protostellar jets, the gas falls directly onto the protostar without forming a disk, and only a single massive star forms.
In addition, a massive binary stellar system appears when $B_{\rm amb} \sim 10^{-12} (n/1\cm)^{-2/3}$\,G.
Therefore, the magnetic field determines the end result of the formation process (cluster, binary or single star) for Population III stars.
Moreover, no persistent circumstellar disk appears around the protostar regardless  of the magnetic field strength, which may influence the further evolution of Population III stars.
%%Although the infalling gas tends to form a circumstellar disk, protostars orbiting the primary star break up the disk.

\end{abstract}
\begin{keywords}
accretion, accretion disks---binaries:general---cosmology:theory---early universe---ISM:magnetic fields---
stars: formation
\end{keywords}

\section{Introduction}
\label{sec:intro}
Stars formed in primordial gas clouds, turned on the light in a dark universe and strongly affected the subsequent evolution of the universe. 
Thus following the dark age, the first stars, the so-called Population III stars, play important roles in the universe.
The first star formed in the first collapsed object (or minihalo) at a redshift of $z\sim10-20$ \citep{haiman96,tegmark97,bromm99,bromm02,abel02,yoshida03,yoshida06,yoshida08}. 
Cosmological simulations have shown that the first collapsed objects had a baryonic mass of $\sim 10^3-10^4\msun$ and a temperature of $\sim200-300\,$K \citep[e.g.,][]{bromm02}.
Until several years ago, it was thought that only a single massive  star with a mass of $\simeq100\msun$ forms in such an object \citep[see the review by][]{bromm04}.
The formation of Population III stars has been studied mainly by numerical simulations because we cannot directly observe them.
To determine the properties of Population III stars, we need to understand the details of their formation. 
%%their formation process of Population III stars in detail.
%%However, it is difficult to confidently determine a typical mass of Population III stars 
%%Thus, a more supplicated calculation is necessary to investigate the properties of Population III stars.
Theoretically, the star formation process can be divided into two phases \citep[e.g.][]{whitworth85}: the gas collapsing (or prestellar) and gas accretion (or protostellar) phases.
The former covers the phase after the cloud begins to collapse, but before the (first) hydrostatic core forms.
%%phase is the period before protostar formation after the nascent cloud begins to collapse.
%%During this phase, the (primordial) gas cloud continues to collapse until the hydrostatic core (i.e. protostar) forms.
The latter phase is the period after protostar formation, during which gas accretes onto the protostar or the region near it.

By the use of one-zone or one-dimensional calculations, Omukai and his collaborators investigated the formation of Population III stars in both the gas collapsing \citep{omukai98,omukai00,omukai05,omukai10}, and gas accretion \citep{omukai01,omukai03} phases. 
They showed that the thermal evolution of the  primordial gas cloud differs considerably from that of present-day clouds  in the gas collapsing phase. 
They also showed that in the gas accretion phase,  protostellar radiation cannot halt mass accretion onto the protostar and a high mass accretion rate lasts a long time. 
Finally, they concluded that both high mass accretion and inefficient radiative feedback from the protostar produce a massive (single) Population III star with a mass of $M\sim100\msun$ in the primordial cloud.

Multidimensional effects such as those of rotation, non-spherically symmetric accretion and magnetic fields are ignored in one-zone and one-dimensional calculations, although they should affect the formation of Population III stars.
In reality,  it is expected that, during the gas accretion phase,  gas accretes onto a proto-Population III star through a circumstellar disk, and does not fall directly onto its surface.
From an analogy to present-day star formation \citep[e.g.][]{tsuribe99}, we can easily imagine that a high-mass accretion rate onto the circumstellar disk induces fragmentation and subsequent multiple star formation.
In addition, even during the gas collapsing phase, a non-spherically symmetric collapse can induce fragmentation and subsequent binary formation \citep{miyama84,matsu97,matsu99,tsuribe06,tsuribe08}.

Using a three dimensional simulation,  \citet{saigo04} investigated the evolution of a collapsing primordial cloud with a simple initial setting and showed fragmentation before protostar formation \citep[see also][]{machida08,machida09a,machida09b}. 
Recently, starting from the cosmological setting, \citet{turk09} also showed the possibility of fragmentation during the gas collapsing phase.
These studies imply that, in a primordial cloud, fragmentation during the gas collapsing phase causes the formation of a (wide) Population III binary.
However, in these studies, the authors investigated only the cloud evolution before protostar formation (i.e. during the gas collapsing phase) because the calculation time step becomes increasingly short as gas density increases.
Thus, we cannot calculate the cloud evolution for a long duration after protostar formation without any artifice.
However, to determine how a Population III star forms in the first collapsed object, we need to calculate the cloud evolution even during the gas accretion phase.

\citet{clark08} first focused on the evolution of a primordial cloud during the gas accretion phase following the gas collapsing phase by using a sink treatment and showed multiple fragmentation in the circumstellar disk after first protostar formation.
Then, by using sink particles or sink cells,  many studies have confirmed multiple fragmentation during the gas accretion phase \citep{stacy10,greif11,clark11a,clark11b,smith11}.
All recent studies indicate that Population III stars are born in star clusters, as observed in present-day star-forming regions.
In addition,  although different sizes of sink radius (or accretion radius) were adopted in these studies, fragmentation always occurred.
This implies that fragmentation occurs on any spatial scale in primordial clouds.
To determine the minimum scale of fragmentation, we have to calculate primordial clouds without sinks.
Very recently, \citet{greif12} investigated  primordial cloud evolution without sinks, and showed that fragmentation can occur even at a considerably small scale comparable to the protostellar radius.
Their results might mean that we have to calculate  cloud evolution without sinks to investigate fragmentation and further evolution of primordial cloud and Population III stars.

In addition to treating sinks differently, these studies differed in numerical settings such as the initial conditions (cosmological setting or equilibrium sphere), thermal treatment (barotropic equation of state or radiative cooling) and chemical networks.
%%Among these studies, beside the different treatment of sink, there are some other differences in numerical settings such as the initial condition (cosmological setting or equilibrium sphere), thermal treatment (barotropic equation of state or radiative cooling) and chemical networks and so on.
Although these differences produced slight quantitative differences in the results, such as the resultant stellar mass, number of fragments and fragmentation scale, fragmentation is never suppressed.
%%This means that fragmentation is controlled by the mass accretion and efficiency of the angular momentum transfer in the disk, not by the initial condition and thermal evolution.
%%Note that the thermal evolution is partly related to the angular momentum transfer because spiral arms tends to develop with effective cooling \citep{toomre64}.
When the mass accretion rate onto the circumstellar disk from the infalling envelope is larger than the mass flow rate to the protostar from the disk, the disk surface density (or disk mass) continues to increase, and the disk becomes gravitationally unstable to fragmentation \citep{clark11a,greif11}. 
Thus, fragmentation inevitably occurs in a primordial cloud which  has a high mass accretion rate \citep[e.g.][]{omukai03,hosokawa09b}.
In addition, \cite{smith11} showed that the accretion luminosity of the primordial protostar does not substantially affect gas accretion.
\cite{greif11} also showed that fragmentation cannot be suppressed even when an unrealistically strong radiation field from primordial protostars is assumed. 
Thus,  accretion luminosity feedback from primordial protostars has a minimal effect on fragmentation or star cluster formation \citep{smith12}.

As a result, there is a consensus that fragmentation is a general occurrence and that multiple Population III stars form in a single primordial cloud.
However, previous studies ignored the effects of the magnetic field.
Although the strength of the magnetic field is still controversial \citep{langer03,ichiki06,xu08,doi11,schober12}, the magnetic field may change the picture of fragmentation in collapsing primordial clouds.
The magnetic field can effectively transfer angular momentum in the circumstellar disk by magnetic braking \citep{mellon08,mellon09,li11}.
\citet{machida08c} showed that the magnetic field suppresses fragmentation in a primordial cloud during the gas collapsing phase.
Recently, \citet{turk12} investigated Population III star formation from realistic initial condition assuming an identical uniform magnetic field and pointed out that circumstellar disk formation tends to be suppressed in a magnetized collapsing primordial cloud during the gas collapsing phase.

The effects of the magnetic field on Population III star formation during the gas collapsing phase were investigated  in a few studies to date, whereas those during the gas accretion phase have not been studied  with multi-dimensional calculations. 
This is because the magnetic field is incompatible with the sink treatment by which we can investigate the gas accretion phase. 
In the sink treatment, the gas inside the sink radius is removed, whereas the magnetic field cannot be removed because of the div$B$=0 constraint. 
Thus, we need some other method to investigate the gas accretion phase with a magnetic field for a long duration.

In this study, following \citet{machida08c}, which investigated the cloud evolution in the gas collapsing phase, we study Population III star formation in weakly magnetized primordial clouds, especially during the gas accretion phase.
First, we model the relationship between protostellar mass and radius with a modified equation of state to realize long term calculation in the gas accretion phase.
This modelling make it possible to treat the magnetic field even during the gas accretion phase without sinks. 
Then, we calculate the evolution of an unmagnetized primordial cloud with a simple initial setting and compare our results with a recent highly developed simulation without sinks \citep{greif12}  to validate our results.

Finally, we calculate the evolution of weakly magnetized clouds and discuss the effect of the magnetic field on the formation of Population III stars, in which large-scale ordered magnetic fields  parallel to the initial rotation axis are assumed.
Note that, however, some recent studies pointed out the possibility of generation of small-scale disordered fields \citep[e.g.][]{schleicher10,sur10,federrath11}, which can affect the disk formation and fragmentation in star forming clouds \citep{commercon11, seifried12}. 
We discuss the initial magnetic configuration and its effects in \S\ref{sec:mag-config}.

This paper is structured as follows.
The framework of our models and the numerical method are described in \S 2, and the numerical results are presented in \S 3. 
We discuss the calculation results, compare the formation of primordial stars with that of present-day stars in \S 4 and summarize our results in \S 5.

\section{Model and Numerical Method}
\label{sec:model}

\subsection{Basic Equations}
\label{sec:basic}
\begin{figure}
\includegraphics[width=150mm]{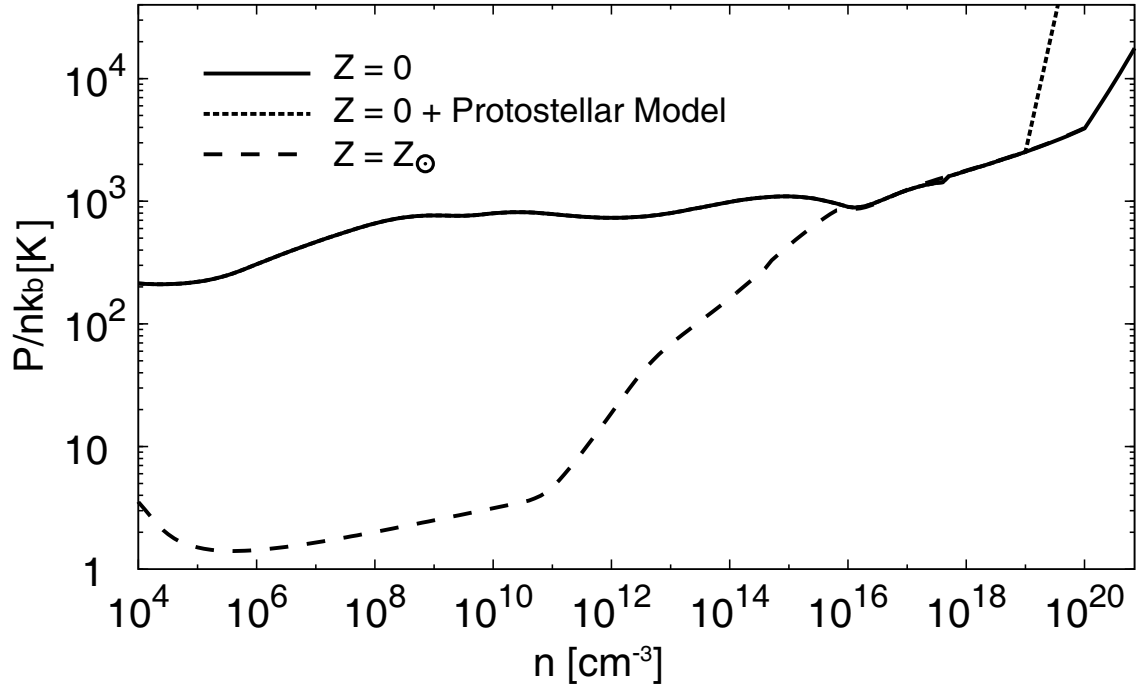}
\caption{
Gas pressure $P$ normalized by number density $n$ for zero-metallicity $Z=0$ gas ({\it solid line}) with our protostellar model ({\it dotted line}) against the hydrogen number density, where $k_{\rm b}$ is the Boltzmann constant. 
Solar metallicity gas ({\it dashed line}) is also plotted.
}
\label{fig:1}
\end{figure}

To calculate the evolution of primordial magnetized clouds  before and after primordial protostar formation, we use three-dimensional resistive magnetohydrodynamics (MHD) equations, including self-gravity:
\begin{eqnarray} 
\label{eq:cont}
& \dfrac{\partial \rho}{\partial t}  + \nabla \cdot (\rho \vect{v}) = 0, & \\
& \rho \dfrac{\partial \vect{v}}{\partial t} 
    + \rho(\vect{v} \cdot \nabla)\vect{v} =
    - \nabla P - \dfrac{1}{4 \pi} \vect{B} \times (\nabla \times \vect{B})
    - \rho \nabla \phi, & \\ 
& \dfrac{\partial \vect{B}}{\partial t} = 
   \nabla \times (\vect{v} \times \vect{B}) + \eta \nabla^2 \vect{B}, & 
\label{eq:reg}\\
& \nabla^2 \phi = 4 \pi G \rho, &
\label{eq:grav}
\end{eqnarray}
where $\rho$, $\vect{v}$, $P$, $\vect{B} $, $\eta$, and $\phi$ denote the density, velocity, pressure, magnetic flux density, resistivity and gravitational potential, respectively.
To self-consistently determine the thermal pressure and magnetic resistivity $\eta$, we first calculated the thermal evolution by using a free-falling one-zone model starting from a hydrogen number density of $n_{\rm H}=0.1\cm$ and a temperature of $T=300$\,K, in which a non-equilibrium chemical reaction network in the primordial gas is solved.
This calculation is basically identical to that of \citet[][primordial case]{omukai05}, although we newly add the species of Li, LiH, Li$^+$, Li$^-$, LiH$^+$, Li$^{++}$, and Li$^{+++}$ to the chemical networks, in addition to H, H$_2$, H$^+$, H$_2^+$, H$_3^+$, H$^-$, He, He$^+$, He$^{++}$, HeH$^+$, D, HD, D$^+$, HD$^+$, D$^-$ and e$^-$.
Although the newly included species do not significantly affect thermal evolution, they play a critical role in estimating the ionization degree and/or magnetic diffusivity \citep[for details, see][]{maki04,doi13}. 
%% in which the CMB radiation at ???K is imposed.
The thermal evolution of the collapsing primordial gas is plotted against the gas number density as a solid line in Figure~\ref{fig:1}.
For reference, the thermal evolution for a solar metallicity gas \citep{doi13} is also plotted by as a dashed line in Figure~\ref{fig:1}.

The thermal evolutions are almost the same as those in \citet{omukai00} and \citet{omukai05}. 
Instead of solving the thermal evolution and chemical reactions in our three-dimensional calculation, we use the barotropic relation, which is denoted in Figure~\ref{fig:1} by a dotted line (for details, see \S\ref{sec:psmodel}).
Although we recognized that the long-term thermal evolution differs significantly from the one-zone model \citep{clark11b}, we use the result of the one-zone model to  focus on the effects of the magnetic field. 
Note that we did not calculate the primordial cloud evolution for a very long duration ($\sim1000$\,yr at best).
Almost the same barotropic relation was used in very recent study \citep{vorobyov13}, in which the long-term evolution ($\sim5\times10^4$\,yr) of the accretion phase of Population III stars was investigated without magnetic field.

\subsection{Resistivity and Magnetic Reynolds Number}
The magnetic field plays an important role in the present-day star formation.
%%The magnetic field is amplified as the cloud collapses, whereas, in present-day clouds, it is 
%%significantly dissipated by Ohmic dissipation in the high-density gas region of 
%%$n\gtrsim 10^{12}\cm$ \citep{nakano02,tassis07a, tassis07b,kunz10}. 
While the magnetic field is amplified as the cloud collapses, in present-day clouds there is significant dissipation via Ohmic diffusion once the gas reaches a density of $n\gtrsim 10^{12}\cm$ \citep{nakano02,tassis07a, tassis07b,kunz10}.  
Thus, we have to consider Ohmic dissipation to investigate the magnetic evolution in present-day clouds \citep{machida07}.
On the other hand, the magnetic dissipation exacerbates the angular momentum problem but promotes disk formation, because it lowers the efficiency of the magnetic braking (\citealt{machida11c}; but, \citealt{krasnopolsky10} claimed that the Ohmic dissipation rarely affects the disk formation).  
Thus, we should investigate the evolution of the present-day cloud with the Ohmic dissipation \citep[e.g.,][]{machida07,machida11c,tomida13}. 
Note that other effects such as the disordered fields \citep{commercon11, seifried12} and turbulent reconnection \citep{Santos-Lima12} possibly promote disk formation in a strongly magnetized cloud (for details, see \S\ref{sec:mag-config}).
Instead, it is considered that the Ohmic dissipation (or magnetic dissipation) is less effective in primordial collapsing clouds due to the high gas temperature and absence of dust grains (see below).

\begin{figure}
\includegraphics[width=150mm]{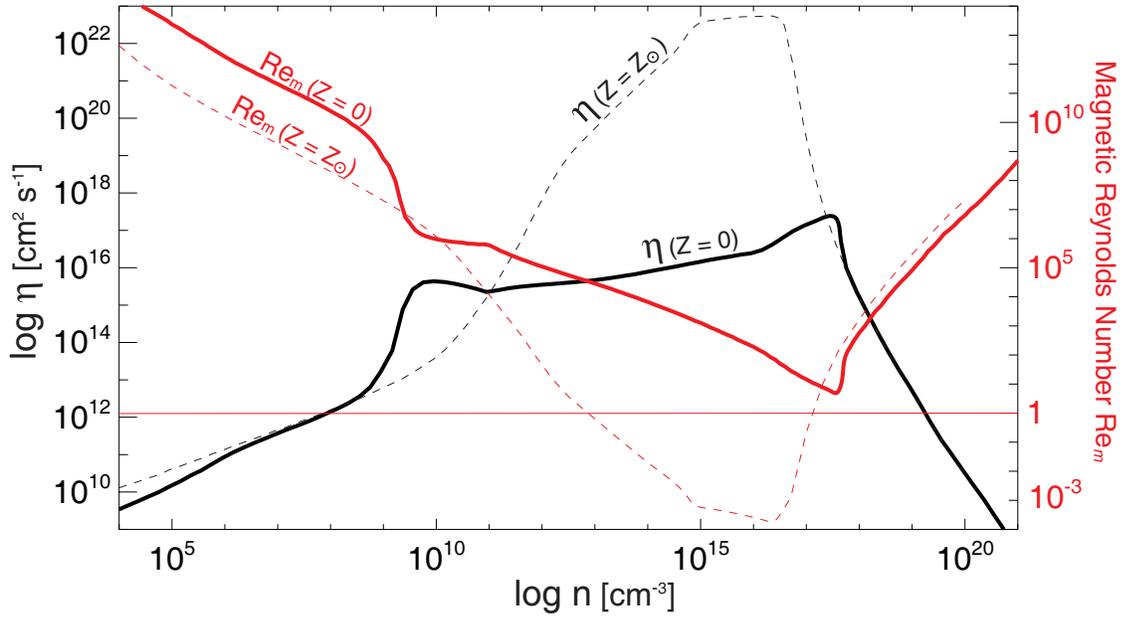}
\caption{
Resistivity $\eta$ ({\it left axis}) and magnetic Reynolds number Re$_{m}$ ({\it right axis}) as functions of number density.
Solid lines denote the zero-metallicity case ($Z=0$);  broken lines denote the solar metallicity case ($Z=Z_\odot$).
}
\label{fig:2}
\end{figure}

With the one-zone calculation, we estimate the ionization degree and resistivity. 
%%The resistivity $\eta$ is related to the ionization degree $X_{\rm e}$ \citep{nakano02,machida07} as 
The resistivity $\eta$ is estimated as 
\begin{equation}
\eta = \dfrac{c^2} { 4 \pi \sigma_c},
%%\eta = \dfrac{740}{X_e} \sqrt{\dfrac{T}{{10\,{\rm K}}}}\ \ \ {\rm cm}^2\,{\rm s}^{-1}.
\end{equation}
where $c$ is the speed of light.
The electrical conductivity $\sigma_c$ is derived using equations (41) and (43) in \citet[][see also \citealt{nakano02,machida07}]{nakano86}. 
The resistivity (black lines) and magnetic Reynolds number (red lines) for primordial (solid lines) and solar (broken lines) metallicity gases are shown in Figure~\ref{fig:2}.
The magnetic Reynolds number in a collapsing cloud can be written as, 
\begin{equation}
Re_{\rm m} \equiv \dfrac{v_{\rm f}\,  \lambda_{\rm J}} {\eta} = \dfrac{c_s^2}{\eta}\, \pi^{3/2}
\left( \dfrac{4}{3\,G\,\rho_{\rm c}} \right)^{1/2},
\label{eq:magR}
\end{equation}
where $c_s$, $G$ and $\rho_c$ are the sound speed, gravitational constant and central density, respectively, 
and $v_{\rm f}$ ($\equiv [(4/3)\pi G\, \lambda_{\rm J} \rhoc]$ and $\lambda_{\rm J}$ ($\equiv [\pi c_s^2/(G\, \rho_c)]^{1/2}$ are the free-fall velocity and Jeans length \citep{machida07}.

Figure~\ref{fig:2} shows that for the solar metallicity gas, the resistivity reaches a peak value at $n\sim10^{15}-10^{16}\cm$, and the magnetic Reynolds number reaches Re$_{\rm m}<1$ in the region  of $10^{13}\cm \lesssim n \lesssim 10^{17}\cm$, where the magnetic field dissipates significantly.
Without the removal of the magnetic field, the solar metallicity cloud cannot collapse further to form a protostar (the so-called magnetic flux problem).
On the other hand, the ionization rate in the primordial gas cloud is considerably higher, and thus the resistivity is lower than those for the solar metallicity gas. 
This is because the primordial gas is hotter than the solar-metallicity gas (Fig.~\ref{fig:1}), and thermal ionization does not significantly lower the ionization rate.
In addition,  free electrons are effectively absorbed by dust grains as gas density increases and ionization rate decreases in the solar-metallicity gas, whereas the primordial gas does not contain dust grains \citep{nakano02,maki04,maki07}. 
As a result, the resistivity of the primordial gas cloud is $\sim 10-10^7$ time higher than that for the solar metallicity gas in the range of $10^{13}\cm \lesssim n \lesssim 10^{17}\cm$, and the magnetic Reynolds number never reaches Re$_{\rm m}<1$ in the primordial cloud \citep[see also][]{maki07}.

However, even in a primordial gas cloud, the magnetic Reynolds number is $Re_{m}\sim1-10$ in the region of $n\sim10^{17}-10^{18}\cm$, where the circumstellar disk is expected to form.
Thus, the magnetic field can dissipate in this region in a primordial cloud.
In addition, the magnetic Reynolds number derived in equation~(\ref{eq:magR}) is not applicable, especially in the gas accretion phase after first protostar formation.
We roughly estimate the magnetic Reynolds number with simple assumptions regarding the freefall velocity and Jeans length. 
However, the rotation enables the infalling gas to remain around the primordial protostar for a longer duration.
Thus, Ohmic dissipation is expected to be more effective in the gas accretion phase.
Because it is difficult to estimate the magnetic Reynolds number in the accretion phase, we need to calculate the dissipation of the magnetic field by resistive MHD calculations.
Adopting the gas pressure and resistivity shown in Figures~\ref{fig:1} and \ref{fig:2}, we calculate the evolution of primordial gas clouds using equations~(\ref{eq:cont})-(\ref{eq:grav}).

Note that the ionization rate (or resistivity) in Figure~\ref{fig:2} was derived using the one-zone calculation in the gas collapsing phase.
Thus, strictly speaking, it is not applicable for the gas accretion phase, in which the disk forms.
However, we adopt it as an approximate indicator of resistivity.
In reality, the magnetic dissipation by the Ohmic dissipation is not significant in {\it primordial} clouds, as shown in \citet{maki04,maki07}.
Therefore, it is considered that a certain difference in the ionization degree does not qualitatively change the results.
Moreover, because we calculate the cloud evolution for a short duration after the first protostar formation, it is expected that the ionization rate in the gas accretion phase does not significantly differ from that in the gas collapsing phase.

\subsection{Initial Conditions}
\label{sec:initial}
Cosmological simulations have shown that the natal clouds for Population III  stars have characteristic temperatures of approximately a few hundred K and  densities of $n\sim10^3-10^4\cm$ and are in an almost  equilibrium state \citep[e.g.][]{abel00,abel02,bromm02,yoshida06}. 
To mimic this, as the initial condition we adopt the critical Bonner-Ebert (BE) sphere, which is characterized by two parameters: the central density and gas temperature.
We set the central density to $n_0=10^4\cm$ and the gas temperature to $T_0 = 214$\,K, which were derived by the one-zone calculation.
With these parameters, the initial cloud  has a radius of $R_{\rm cl} = 2.8\times10^5$\,AU (1.4\,pc).
We enhance the density by a factor of $1.4$ to promote contraction, and the initial cloud has a mass of $M_{\rm cl}=811\msun$. 
In addition, to break the axisymmetry, non-axisymmetric density perturbations of $m=2$ ($10\%$) and 3 ($1\%$) are added to the initial state.

Although we initially assume a spherical cloud with characteristics similar to those seen in cosmological simulations, these initial conditions can be assumed to influence the results.
However, the initial cloud properties except for the physical quantities at the centre of the cloud, affect the subsequent cloud evolution very little.
After the cloud begins to contract, the gas undergoes a self-similar solution \citep{larson69,omukai98}, in which the density profile is uniquely determined, independent of the initial cloud conditions \citep{larson03}.
Thus, the initial cloud conditions, such as density distribution and cloud shape, are not crucial to investigating the collapsing phase of primordial clouds and  the early phase of protostellar evolution.
This is natural because the evolution timescale of a cloud is given by the {\em local} free-fall timescale, which is proportional to $\propto \rho^{-1/2}$.
Thus, only a central high-density cloud region first collapses (or evolves), whereas the outer low-density region is left behind from the cloud evolution.
Therefore, the outer cloud region never affects the early phase of first protostar formation, which is the focus of this study.
Note that the outer low-density gas falls onto the protostellar system and affects the protostellar evolution with time.
Note also that, however,  we cannot calculate the cloud evolution for the long time period until the outer gas falls onto the centre of the cloud because we resolve the protostellar radius without sink cells (see \S\ref{sec:psmodel}).

\citet{machida08} investigated the evolution of magnetized primordial clouds and pointed out that only the magnetic field strength and rotation rate at the centre of the cloud determine the early evolution of protostars.
Thus, we also parameterize them in this study.
In each model, the cloud rotates rigidly around the $z$-axis and a uniform magnetic field parallel to the $z$-axis is adopted in the entire  computational domain.
Using a combination of two parameters, the magnetic field strength ($B_0$)  and rotation rate ($\Omega_0$), we construct 14 models.
They are  listed in Table~\ref{table:1}, which also lists the magnetic $\gamma_0$ ($\equiv E_{\rm mag}/E_{\rm grav}$) and rotational $\beta_0$ ($\equiv E_{\rm rot}/E_{\rm grav}$) energies normalized by gravitational energy, where $E_{\rm mag}$, $E_{\rm rot}$ and $E_{\rm grav}$ are the magnetic, rotational and gravitational energies of the initial cloud, respectively.
In addition, we estimate the mass-to-flux ratio normalized by the critical value
\begin{equation}
\mu = \left( \dfrac{M}{\Phi} \right) / \left( \dfrac{M}{\Phi} \right)_{\rm cri}, 
\end{equation}
where $(M/\Phi)_{\rm cri} = 485$ is the critical mass-to-flux ratio \citep{mouschovias76,tomisaka88a,tomisaka88b}.
The normalized mass-to-flux ratio for each model is also listed in Table~\ref{table:1}.

%%As described in Table~\ref{table:1}, 
We adopted the magnetic field strengths of $0 <B_0<10^{-5}\,{\rm G}$; models 1, 2 and 14 have no magnetic field (unmagnetized models).
%% and we call this model unmagnetized model.
Because the strength of the magnetic field in primordial clouds is highly controversial, we adopt a wide range of magnetic field strengths.
Although it is also difficult to determine the rotation rate (or rotational energy) in primordial clouds, there are some indications. 
Observations show that nearby molecular cloud cores have $10^{-4}<\beta_0 < 1.4$  with a typical value of $\beta_0 \sim0.02$ \citep{goodman93,caselli02}. 
Cosmological simulations showed that the first-star-forming cores have $\beta\lesssim 0.1$ \citep{bromm02,yoshida06}.
Thus, the natal cloud of first stars are expected to have a non-negligible rotation rate.
To limit the number of models, we adopt plausible values of the rotation rate, $\Omega_0 =  8.5\times10^{-15}$\,s$^{-1}$, $2.7\times10^{-15}$\,s$^{-1}$ and  $8.5\times10^{-16}$\,s$^{-1}$, which correspond to $\beta_0=10^{-2}$, $10^{-3}$ and $10^{-4}$, respectively.

The maximum value of $\beta=10^{-2}$ adopted in this study roughly corresponds to the mean value of nearby molecular cloud cores, and is comparable to or somewhat smaller than that derived by cosmological simulations. 
In principle, it would be advantageous to study higher $\beta$ parameters, however such conditions are, at present, computationally challenging. 
As such, we limit the focus of this present study to the effect of the magnetic field in the presence of somewhat smaller initial rotational energies.
%%Although we may have to adopt the initial cloud having a lager $\beta$, it is difficult to calculate the evolution of such a cloud with our numerical method (\S\ref{sec:method}). 
%%In addition, since it requires a huge CPU time to calculate the Population III star formation in a primordial cloud, we have to limit the number of calculations.
%%In this study, to focus on  the effect of the magnetic field on the Population III star formation, we adopt somewhat smaller initial rotational energies. 
To obtain the general picture of the Population III star formation, we have to calculate it in a wider parameter range or in various minihalos in future.
%%Note that rotation rates adopted in this study may be somewhat smaller than the typical value.
%%Note also that we should investigate fragmentation process with preferably smaller rotation rate since we can conclude that fragmentation generally occurs if fragmentation occurs even with considerably smaller rotation rate.

\subsection{Numerical Method}
\label{sec:method}
To calculate spatial scales which differ considerably, we use the nested grid method  \citep[for details, see][]{machida04,machida05a,machida05b}.
Each level of a rectangular grid has the same number of cells, (i, j, k) = (256, 256, 32), in which the cell width ($l$) depends  on the grid level $l$.
The cell width is halved with every increment of the grid level.
The highest level of the grid changes dynamically to ensure the Truelove condition \citep{truelove97}, in which the local Jeans length is resolved at least for eight cells. 
With this spatial resolution, we can properly investigate the fragmentation process in the collapsing cloud.
%%However, we need a more higher spatial resolution to investigate the turbulence and turbulent 
%%dynamo generation, in which the Jeans length should be resolved at least 30 cells 
%%\citep{federrath11}.
Note that while sufficient for resolving the fragmentation process in these systems, our current resolution is not sufficient for resolving the turbulent dynamo, and thus by implication, the turbulent cascade \citep{federrath11}. 

We first prepare five grid levels:  the initial cloud (critical BE sphere) is immersed in the fifth level.
We adopt a computational boundary 32 times larger than the initial cloud radius to safely calculate the evolution of a magnetized cloud.
This treatment can prevent the reflection of Alfv$\acute{\rm e}$n waves at the computational boundary (for details, see \citealt{machida11c,machida12}).
The fifth level of grid has a box size of $L(5)=5.70\times10^5$\,AU (2.76\,pc) with a cell width of $h(5)=223$\,AU, whereas the first level of grid has $L(1)=9.12\times10^6$\,AU (44.16\,pc) and $h(1)=3.56\times10^4$\,AU.
The maximum level of grid is set to $l=21$ with  $L(21)=8.78$\,AU and $h(21)=0.034$\,AU.
Thus, we can cover the structures of both the natal cloud ($\sim$\, pc) and the protostar ($<0.1$\,AU).

After the first protostar forms or the $l=21$ grid is generated, we stop the generation of new finer grid levels.
However, we confirmed that Truelove condition was always fulfilled in both $l=20$ and $21$ grids (see, \S\ref{sec:psmodel}).
When fragmentation occurs, several fragments are ejected from the $l=20$ grid by gravitational interaction long after first protostar formation, and the Truelove condition is violated in such fragments, which are located in the $l\le 19$ grid. 
Note that $l=20$ has a box size of $L(20)=17.6$\,AU. 
We do not trace fragments after they are ejected from the central region or $l=20$ grid (see \S\ref{sec:unmanetized}). 
In addition, such small fragments are not expected to  significantly affect the dynamical evolution and fragmentation around the centre of the cloud.

In this study, finer rectangular grids are fixed to the center of  the collapsing cloud (or the center of coarser grids). 
Thus, we  cannot trace a long-term evolution of the disk because the  high-density gas region that corresponds to the disk expands with  time and the Jeans condition is finally violated in a coarser grid. 
Note that the disk size progressively increases because the gas  accreted later has a larger angular momentum. For this study,  however, the disk and fragmentation region were limited only in l=20  and 21 grids and the Jeans condition was fulfilled in these grids by the end of the calculation, and thus we could safely calculate the  disk evolution and  fragmentation. 
To calculate the disk evolution  for a longer time, we require another technique such as adaptive  mesh refinement.

\subsection{Protostellar Model}
\label{sec:psmodel}
A Population III star gradually increases its radius in the early evolutionary stage.
The radius of proto-Population III star is $R_* \sim1-10\rsun$ at its formation and increases as the protostellar mass increases because the high mass accretion rate with high entropy causes the protostar to swell.
\citet{omukai03} calculated the protostellar evolution with a constant accretion rate of $\dot{M} = 4.41\times10^{-3}\msun$\,yr$^{-1}$.
They showed that the protostar has a radius of $R_* \sim 10\rsun$ when it has a mass of $M_* = 0.01\msun$ and the protostellar radius increases roughly as $R_* \propto M_*^{1/3}$ (the adiabatic accretion phase; see also \citealt{hosokawa09a}, \citealt{smith11}, \citealt{smith12}).
The protostar has a radius of $R_*\sim200\rsun$ when the protostellar mass reaches $M_* \simeq 10-20\msun$.
Then, for $M_*>20\msun$, the protostar enters in the Kelvin-Helmholtz contraction phase and begins to contract.

To investigate the circumstellar disk formation and fragmentation process in multi-dimensional calculations, the sink cell approach is often used to realize a long-term calculation during which the sink radius (or accretion radius) is usually fixed. 
In present-day star formation, the (rotating) first adiabatic core forms with a size of $\gtrsim 10$\,AU \citep{saigo06} before protostar formation.
After protostar formation, the first adiabatic core becomes the circumstellar disk \citep{bate98,bate11,machida10a}, and fragmentation occurs at the scale of the first core ($>10$\,AU).
As a result,  the first adiabatic core gives the typical scale of disk formation and fragmentation.
In addition, during the main accretion stage, the protostellar radius remains almost constant at $\sim \rsun$ \citep[or $<10\rsun$;][]{stahler80,baraffe09,hosokawa11b} and is deeply embedded in the disk (or the remnant of the first adiabatic core), which  has  a size of $\gg 1$\,AU \citep{machida10a,tomida13}.
Therefore, in present-day star formation, we need a spatial scale of $\sim1-10$\,AU to investigate  disk formation and fragmentation.
In other words, we can replace the region inside $r<1$\,AU with a sink.

In contrast, in a primordial collapsing cloud, there is  no typical spatial scale.
As seen in Figure~\ref{fig:1}, before protostar formation ($n\gtrsim 10^{18}\cm$), the gas temperature gradually increases with a polytropic index $\gamma\simeq1.1$ for the primordial cloud \citep{omukai98}, whereas it suddenly increases at $n\sim10^{11}\cm$ and gives the typical scale for the solar metallicity cloud \citep{omukai00,omukai05,omukai10}. 
Therefore, unlike the solar metallicity case, no adiabatic core forms before protostar formation in primordial collapsing cloud.
Thus, it is expected that the circumstellar disk gradually grows, and fragmentation occurs near the protostar.
Note that in the solar metallicity case, a disk-like structure (i.e. the first adiabatic core) with a scale of $>10$\,AU exists even before  protostar formation \citep{bate98,bate11}. 

In addition, as described above, the protostellar radius changes drastically from $\sim10\rsun$ to $\sim 1$\,AU in the range of $0.01\msun \lesssim M_* \lesssim  10\msun$.
Thus, as the protostar grows, it may swallow the circumstellar disk or fragments.
In any case, to study the disk formation and fragmentation processes, we need to resolve the region near the protostar and protostar itself (or protostellar radius) without a sink.
To investigate the effect of the sink, we calculated the evolution of rotating or non-rotating primordial gas clouds with different sink radii in advance (see the Appendix) and confirmed that fragmentation always occurs with a scale several times larger than the sink radius (or the accretion radius).
This indicates that the sink radius artificially yields the fragmentation scale and controls the evolution after protostar formation (or after sink creation).
Note that when protostellar feedback is taken into account with radiative transfer, fragmentation scale is considered to be controlled by it \citep{clark11a,smith11}. 
Thus, the fragmentation scale without protostellar feedback is expected to differ from that with it.

Moreover, we cannot calculate a magnetized cloud without effective magnetic dissipation, as described in \S\ref{sec:intro}.
In summary, we have to calculate the cloud evolution without a sink, resolving the protostellar radius.

However, when we resolve the inner structure of the protostar, we cannot calculate the cloud evolution for a long-duration after protostar formation.
This is because the calculation time step becomes extremely small when resolving the protostar, as seen in \citet{greif12}.
To realize a long-term calculation after protostar formation, we construct a protostellar model in which the protostellar radius is related  to the protostellar mass.
As described in \citet{omukai03}, the protostellar mass can be related to the protostellar radius when the mass accretion rate is given.
First, to estimate the mass accretion rate, we prepared a non-rotating primordial cloud (model 1; see Table~\ref{table:1})  and calculated its evolution with a sink, where the accretion radius is $r_{\rm acc}=0.04$\,AU is adopted (model S1; for details, see Appendix).
Figure~\ref{fig:3} plots the mass accretion rate against the protostellar mass, in which the mass accretion rate is calculated with the accretion gas onto the sink.
This figure shows that the mass accretion rate is in the range of $0.001 \lesssim \dot{M}/(\msun\,{\rm yr}^{-1}) < 0.01$ with an average of $\dot{M} = 5.1 \times 10^{-3} \msun\,{\rm yr}^{-1}$ for $M_* \lesssim 10\msun$.

Then we calculated the evolution of the same cloud (model 1) without a sink, but changed the thermal evolution (or gas pressure) in the range of  $n \gtrsim 18\cm$.
We used the thermal evolution derived by the one-zone calculation described in \S\ref{sec:basic} for $n<10^{18}\cm$, whereas we adopted an artificially constructed polytropic relation for $n>10^{18}\cm$.
We constructed 30 different polytropic relations and checked the relationship between the protostellar mass and radius for each one.
(We calculated an evolution of non-rotating primordial cloud 30 times with different polytropic relations.)
In each relation, the cloud collapse halts at $n\sim10^{18}-10^{20}\cm$, and an adiabatic core forms because we adopted a harder equation of state than that derived by the one-zone calculation.
Then, with this harder equation of state, the adiabatic core (or protostar) swells as the mass increases because the central region cannot collapse further.
Among the 30 polytropic relations (or 30 different calculations), we choose the most plausible one which well describes the mass-radius relation for the primordial protostar shown in \cite{omukai03}.
 Note that the equation of state adopted in this study is physically required, while a stiff equation of state was used in previous studies to halt the cloud contraction or to avoid an extremely short timestep \citep[e.g.][]{bate95,tomisaka02}.

The chosen polytropic relation is plotted as a dotted line in Figure~\ref{fig:1}, and roughly obeys the relation $P\propto \rho^4$ for $n\gtrsim 10^{18}\cm$.
The protostellar radius with the chosen polytropic relation is plotted in Figure~\ref{fig:3}.
In addition, the thin lines in Figure~\ref{fig:3} are the radius-mass relations in \citet{omukai03}, where the protostellar radii were derived with the mass accretion rate of $\dot{M}=4.41\times10^{-3}\mdot$ (solid line) and $8.82\times10^{-3}\mdot$ (dotted line).
Figure~\ref{fig:3} indicates that our protostellar model agrees well with the results of \citet{omukai03}. 
Note that in our calculation,  the mass accretion rate averaged over the range of $M_* \lesssim 10\msun$ is  $\dot{M} = 5.1 \times 10^{-3} \msun\,{\rm yr}^{-1}$.

\begin{figure}
\includegraphics[width=150mm]{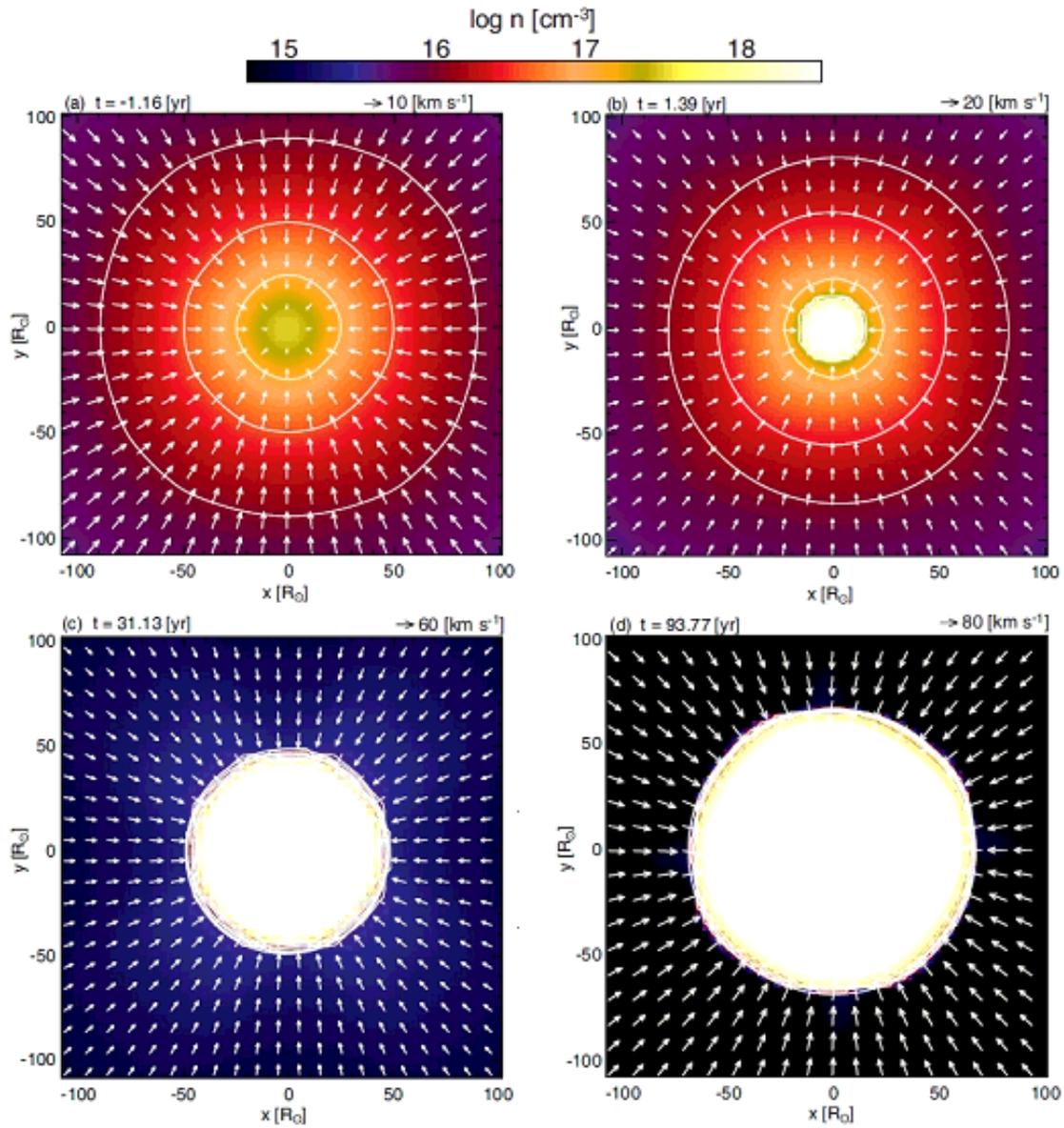}
\caption{
Density (colour and contours) and velocity (arrows) distributions on the equatorial plane around the protostar at different epochs for model 1.
Elapsed time after protostar formation and velocity scale are shown in each panel.
}
\label{fig:4}
\end{figure}

Figure~\ref{fig:4} shows the density and velocity distributions for model 1 at different epochs, where the cloud evolution is calculated with our protostellar model without a sink.
Figure~\ref{fig:4}{\it a} shows the structure of the central cloud region 1.16\,yr before protostar formation, and Figure~\ref{fig:4}{\it b} shows that just after protostar formation ($t_{\rm c}=1.39$\,yr). 
Note that in this paper, we use two different symbols $t$, which indicates the time elapsed after the cloud begins to collapse and 
and $\tc$, which indicates that after protostar formation.
%%: the former $t$ means the elapsed time after the cloud begins to collapse and the latter $t_{\rm c}$ means that 
Note also that we define the protostar formation epoch $t_{\rm c}=0$, at which the maximum density of the collapsing cloud exceeds $n>10^{18}\cm$.
The central white region in Figure~\ref{fig:4}{\it b} - {\it d} corresponds to the protostar.
With our polytropic relation, the protostar has a radius of $\sim10\rsun$ at its formation.
Then the protostar, enclosed by the shock front, swells with time, as seen in Figure~\ref{fig:4}{\it b} - {\it d}. 
In addition, the figure indicates that the gas falls radially onto the protostar without azimuthal (and celestial) motion because the non-rotating cloud is adopted as the initial state.

\begin{figure}
\includegraphics[width=150mm]{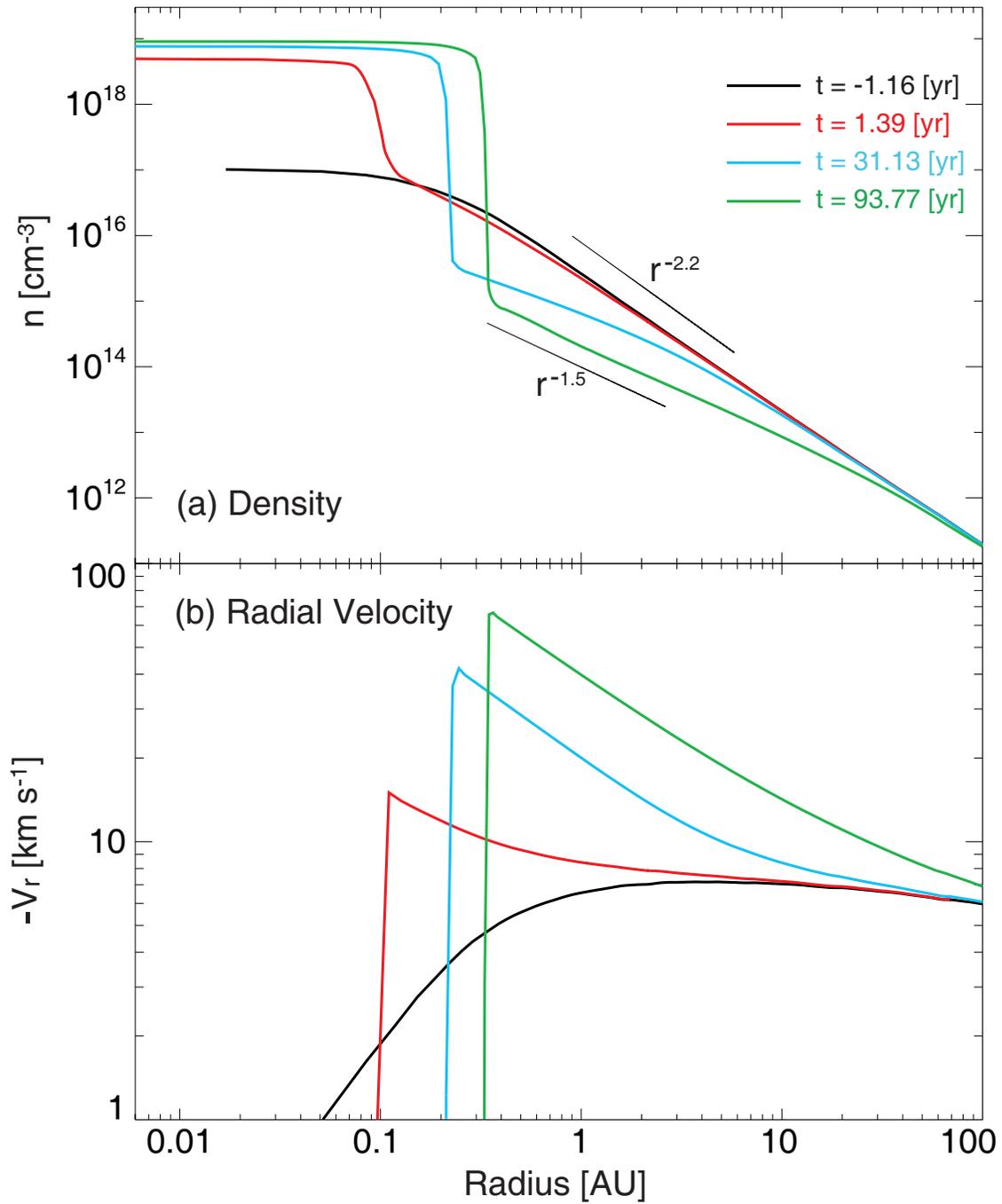}
\caption{
Density (upper panel) and velocity (lower panel) distributions against the radius from the centre of the cloud at four different epochs for model 1.
The relations $\rho\propto r^{-2.2}$ and $\rho\propto r^{-1.5}$ are also plotted in the upper panel.
}
\label{fig:5}
\end{figure}

Figure~\ref{fig:5} shows the radial distribution of the density (upper panel) and velocity (lower panel) at the same epochs as in Figure~\ref{fig:4}.
The sharp increase in the density and velocity (i.e. the shock front) corresponds to the surface of the protostar.
Before protostar formation, the density is proportional to  $\rho \propto r^0$ around the centre of the collapsing cloud, whereas $\rho \propto r^{-2.2}$ in the outer region.
In addition, the radial velocity gradually increases with decreasing radius.
These features are explained well by the self-similar solution in the collapsing cloud \citep{larson69,omukai98}. 
In contrast, after protostar formation, the velocity continues to increase with decreasing radius and suddenly becomes $-v_r=0$ at the protostellar surface (Fig.\ref{fig:5}, lower panel).
The density is proportional to $\rho \propto r^{-1.5}$ near the protostar, whereas it is still proportional to $\rho \propto r^{-2.2}$ far from the protostar.
As described in \citet{whitworth85}, the density profile around the protostar changes from $\rho \propto r^{-2}$ (or $\rho \propto r^{-2.2}$ for a primordial cloud; see \citealt{omukai98}) to $\rho \propto r^{-1.5}$ after protostar formation \citep{larson69, shu77,hunter77}.
Thus, our results agree well with analytical solutions.

By calculating a non-rotating cloud using a polytropic relation, we confirmed that both the protostellar mass-radius relation and the outer density profiles agree well with analytical solutions and one-zone (or one-dimensional) calculations derived in previous studies.
With this protostellar model, we can plausibly treat the tidal interaction and merger between fragments  (\S\ref{sec:unmanetized}). 
%%when fragmentation occurs and some fragments appear in a rotating cloud
However, we do not think that our model can completely reproduce the evolution of the protostar and outer envelope.
We ignored the effects of rotation and the magnetic field  when we constructed the protostellar model (or polytropic relation).
Both rotation and the magnetic field should have little effect on the protostellar evolution because the thermal and gravitational energy dominate the magnetic and rotational energy inside the protostar.
Note that although both cloud rotation and the magnetic field have little effect on the density distribution in a collapsing (primordial) cloud \citep{matsu97,nakamura99,saigo00,machida08,machida09a}, they may change the protostellar radius to some extent.
In addition, we also ignored protostellar feedback, which can affect the infalling envelope around the protostar.
Note that the heating effect of the central protostar on the ambient gas may be ignored because the ambient primordial gas has a high temperature before protostar formation \citep{omukai10}.
Moreover, the protostellar mass-radius relation for $M_* \gtrsim 20\msun$ is not reproduced by our protostellar model, because our model ignored the Kelvin-Helmholtz contraction phase.
It seems difficult to consider these effects when investigating the evolution of a primordial gas cloud.
In this study, using our simple protostellar model, we investigate the cloud evolution and the impact of the magnetic field on it. 
Our modelling may not be sufficiently adequate for the quantitative estimation of certain characteristics, such as protostellar mass and number of fragments. 
Thus, in this study, we focus mainly on investigating the effects of the magnetic field on the evolution of the primordial cloud during the main accretion phase.
We believe that this study is an important step in understanding the effects of the magnetic field on Population III star formation.

\section{Results}
\label{sec:results}
We calculated the evolution of a primordial cloud with rotation until several hundreds of years after first protostar formation with and without magnetic fields, using the models listed in Table~\ref{fig:1}.
In this section, after we describe the cloud evolution for the model without a magnetic field (model 2) in \S\ref{sec:unmanetized},  we present the models with a magnetic field  in \S\ref{sec:magnetized}.

\subsection{Evolution of Unmagnetized Primordial Cloud}
\label{sec:unmanetized}
\begin{figure}
\includegraphics[width=150mm]{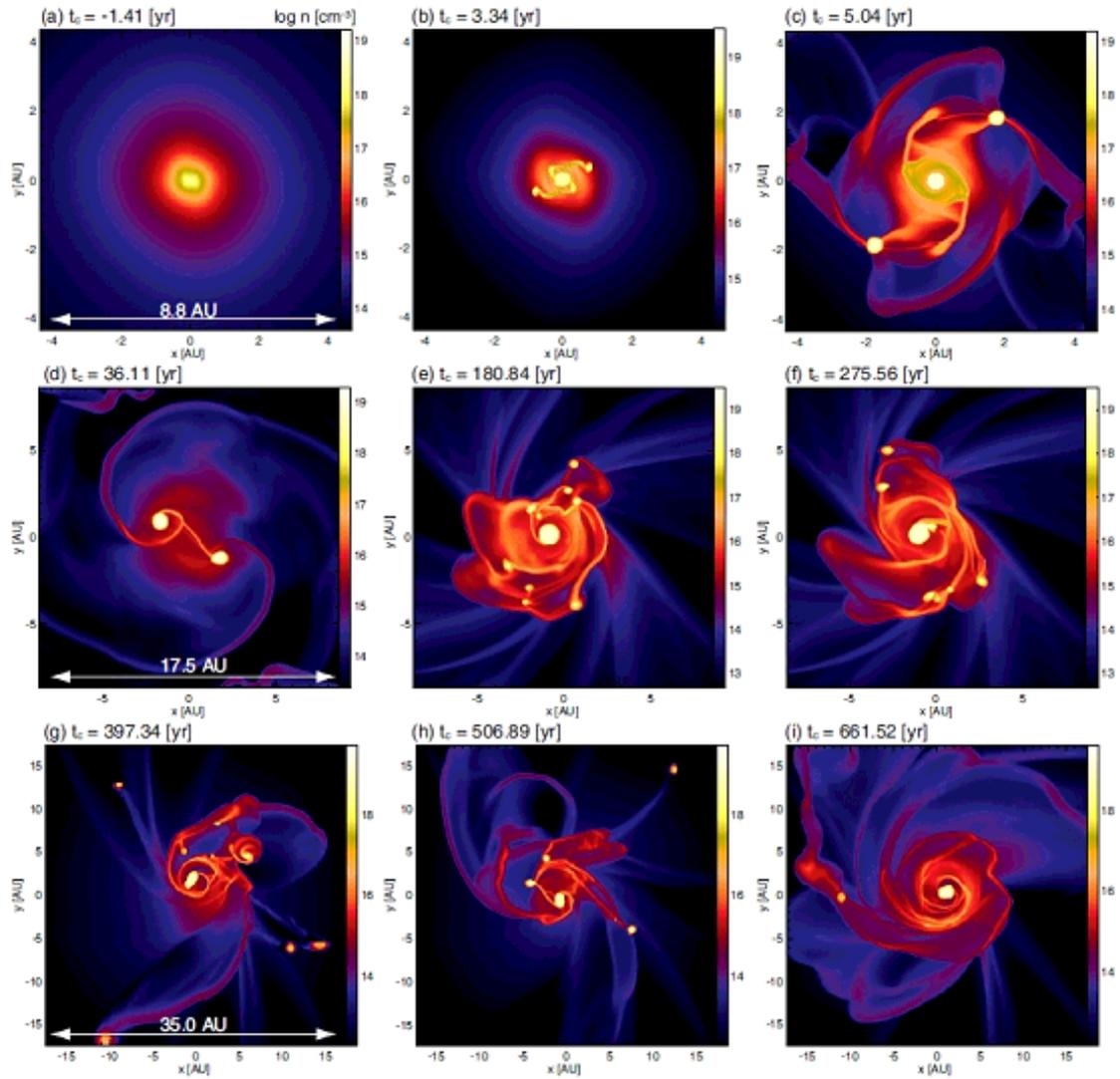}
\caption{
Time sequence images around the centre of the cloud for model 2.
In each panel, the density (colour) distribution on the $z=0$ plane is plotted.
The box size is different in each row.
Elapsed time after protostar formation is given in each panel.
}
\label{fig:6}
\end{figure}

\begin{figure}
\includegraphics[width=150mm]{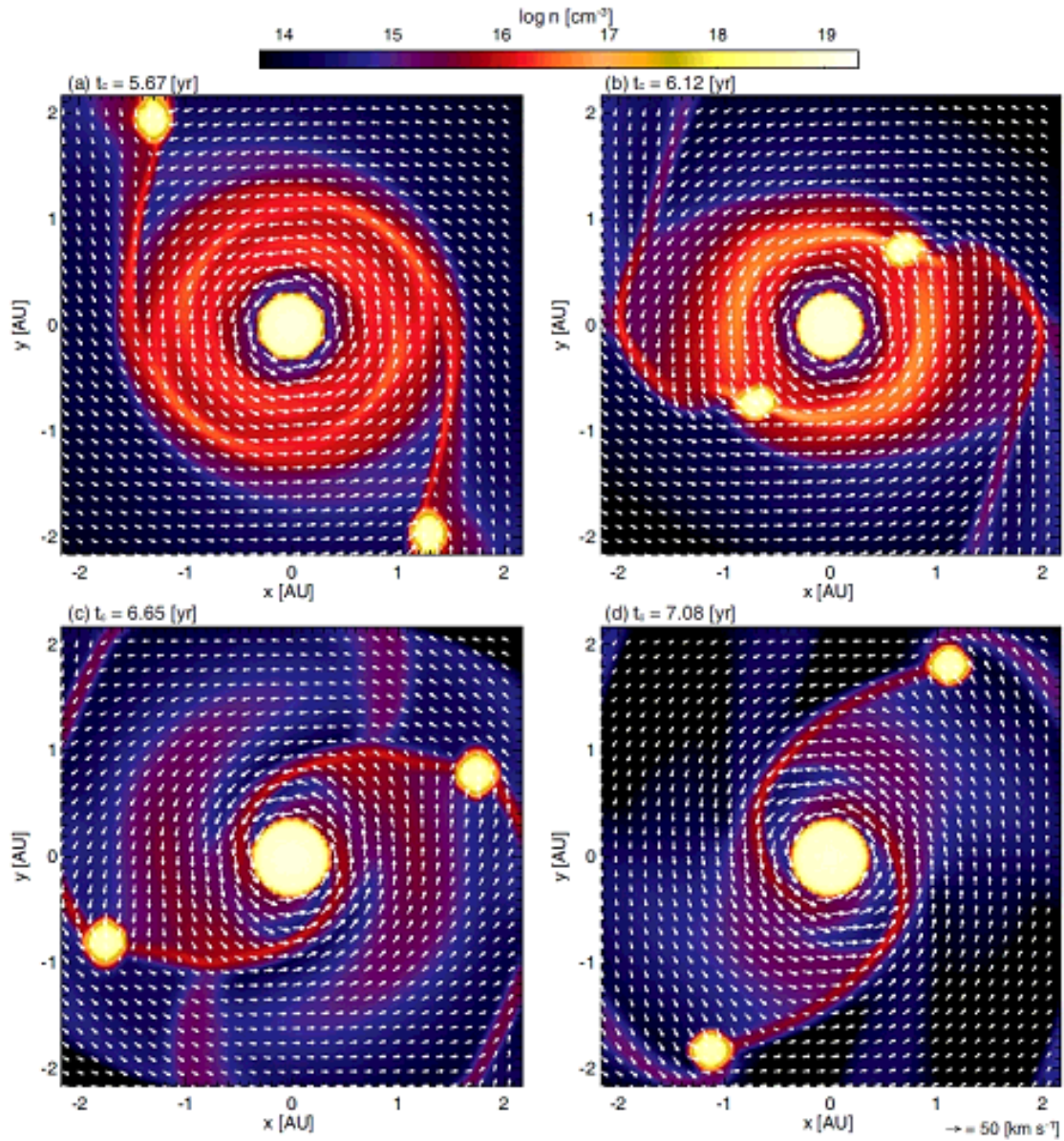}
\caption{
Density (colour) and velocity (arrows) distributions around the centre of the cloud on the $z=0$ plane during $t_c=5.67$\,yr to $7.08$\,yr.
}
\label{fig:7}
\end{figure}

Figure~\ref{fig:6} shows the density distribution around the centre of the cloud at different epochs for model 2.
The initial cloud for model 2 rotates rigidly with an angular velocity of $\Omega_0=8.5\times10^{-16}$\,s$^{-1}$ but is not magnetized ($B_0=0$\,G).
In Figure~\ref{fig:6}{\it a}, a centrally condensed density profile is realized before  protostar formation.
The protostar forms $t=5.563\times10^5$\,yr after the initial cloud begins to collapse.
Just after protostar formation, fragmentation occurs at $\sim1$\,AU far from the first formed protostar (hereafter, the primary star) and two clumps (two fragments) appear, as seen in Figures~\ref{fig:6}{\it b} and {\it c}.
The disk-like structure is also confirmed near the primary star in Figure~\ref{fig:6}{\it c}.
The disk is disturbed by the clumps, which suppress the formation of a (stable or persistent) rotating disk around the primary star.
Figure~\ref{fig:7} shows a close-up view of the density distribution around the primary star during $t_c=5.67$\,yr to $7.08$\,yr.
The density and velocity distributions in Figure~\ref{fig:7}{\it a} indicate that a rotating disk exists in the region of $r \lesssim 1$\,AU.
At this epoch, the clumps orbit in the outer disk region.
Then, their orbits gradually shrink, and  the clumps accumulate disk mass as they approach the primary star (Fig.~\ref{fig:7}{\it b} and {\it c}).
Finally, the clumps acquire a great deal of mass from the disk, and the disk mass is depleted (Fig.~\ref{fig:7}{\it d}).
%%As a result, no persistent (rotating) disk appears around the primary star .

After the first fragmentation event, fragmentation frequently occurs around the primary star, as seen in Figure~\ref{fig:6}{\it e} and {\it f}.
Because the mass accretion rate is quite high in the collapsing primordial cloud, the rotating disk promptly becomes massive and gravitationally unstable.
Therefore, fragmentation occurs, and many clumps appear around the primary star.
In other words, no stable rotating disk, as conceived in present-day low-mass star formation, appears in a primordial cloud, at least during the early stage of star formation.
Moreover, as shown in Figure~\ref{fig:7}, because  clumps  acquire their mass from circumstellar material after they  form, the circumstellar material (or disk) is rapidly depleted.
In summary, immediately after the disk-like structure appears, fragmentation naturally occurs with a high mass accretion rate, and the fragments wipe out the disk.

\begin{figure}
\includegraphics[width=150mm]{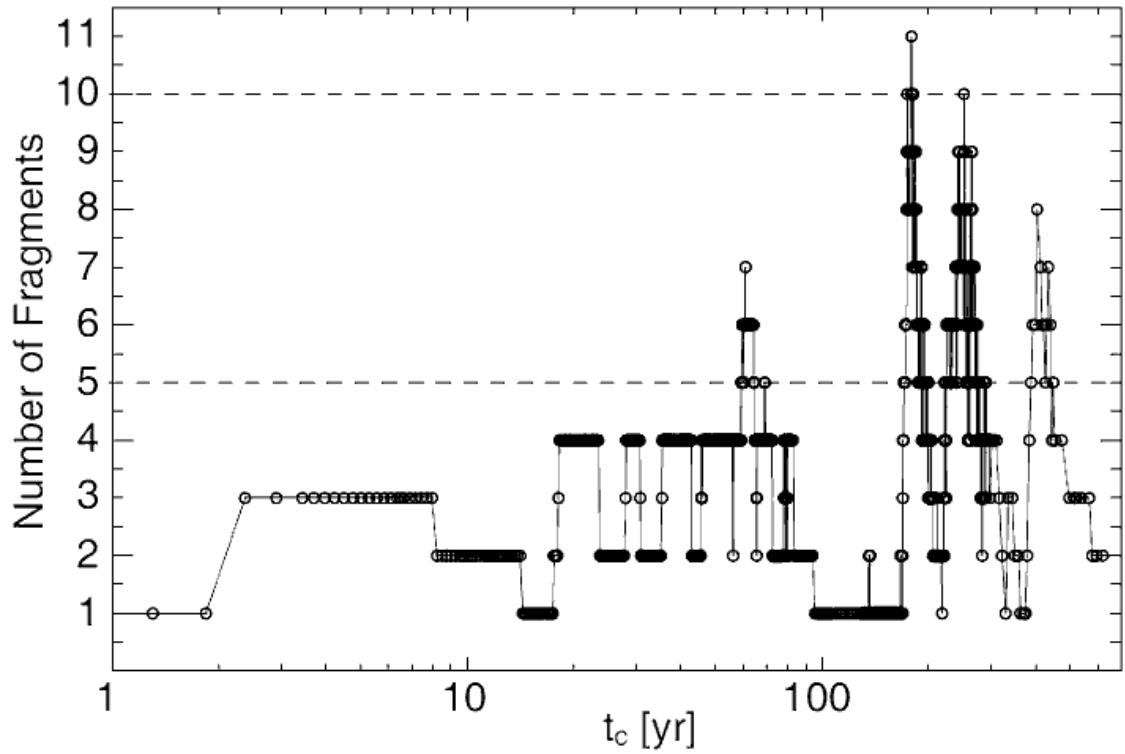}
\caption{
Number of clumps in the region of $r\le10$\,AU against the elapsed time after protostar formation.
}
\label{fig:8}
\end{figure}

The number of clumps is plotted against the time after protostar formation in Figure~\ref{fig:8}, in which the primary star is also counted in the number. 
We identified clumps as isolated fragments having a maximum density of $n>10^{18}\cm$.
Note that we counted the number of clumps only in the region of $r<10$\,AU because we cannot resolve clumps with sufficient spatial resolution in the region of $r\gg10$\,AU (\S\ref{sec:method}).
Note also that we regard the clumps which escaped from the region of $r<10$\,AU as ejected clumps.
The figure shows that fragmentation occurs and two clumps form at $t_c\sim2$\,yr after primary star formation.
Two clumps merges to form a single clump which falls onto the primary star at $t_c=17.1$\,yr.
Then, fragmentation occurs around the primary star, and 2-7 clumps appear during $t_c<200$\,yr.
During this epoch, three clumps are ejected from the region of $r>10$\,AU.
For $\tc>200$\,yr, a maximum of 11 clumps  appear in the region of $r<10$\,AU.
By the end of the calculation, 9 clumps have  escaped from the region of $r<10$\,AU.
In addition, some of the clumps merged with other clumps or the primary star.

\begin{figure}
\includegraphics[width=150mm]{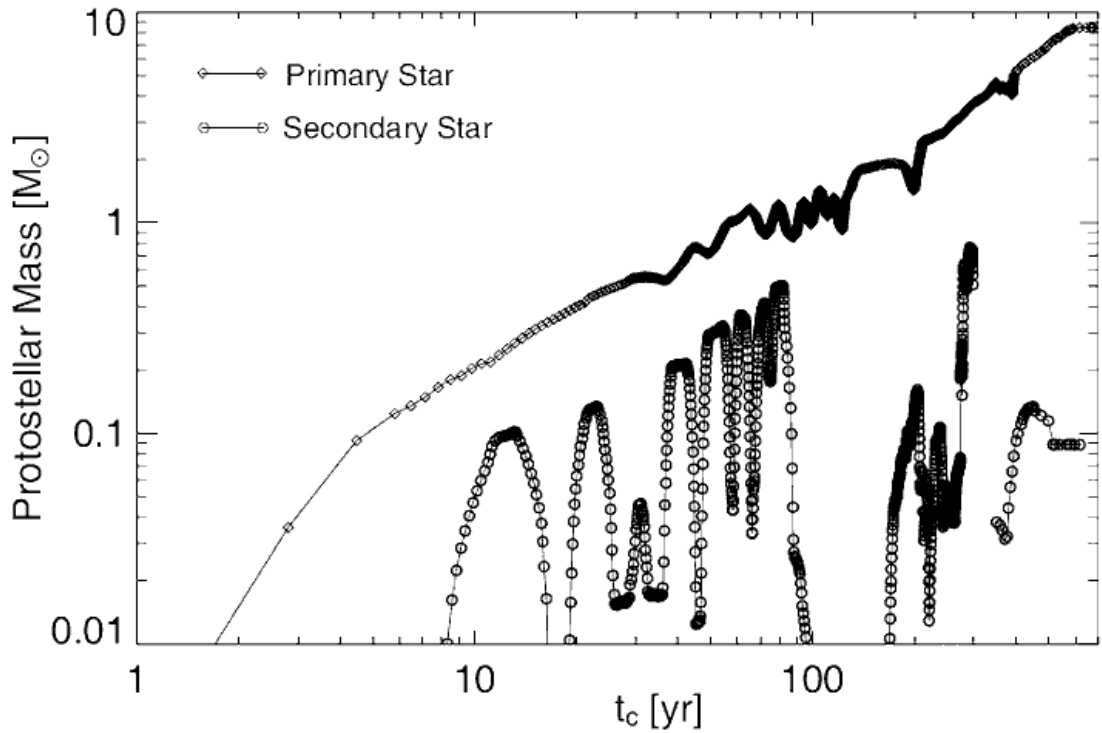}
\caption{
Mass evolution of primary and secondary stars against elapsed time after protostar formation.
}
\label{fig:9}
\end{figure}

\begin{figure}
\includegraphics[width=150mm]{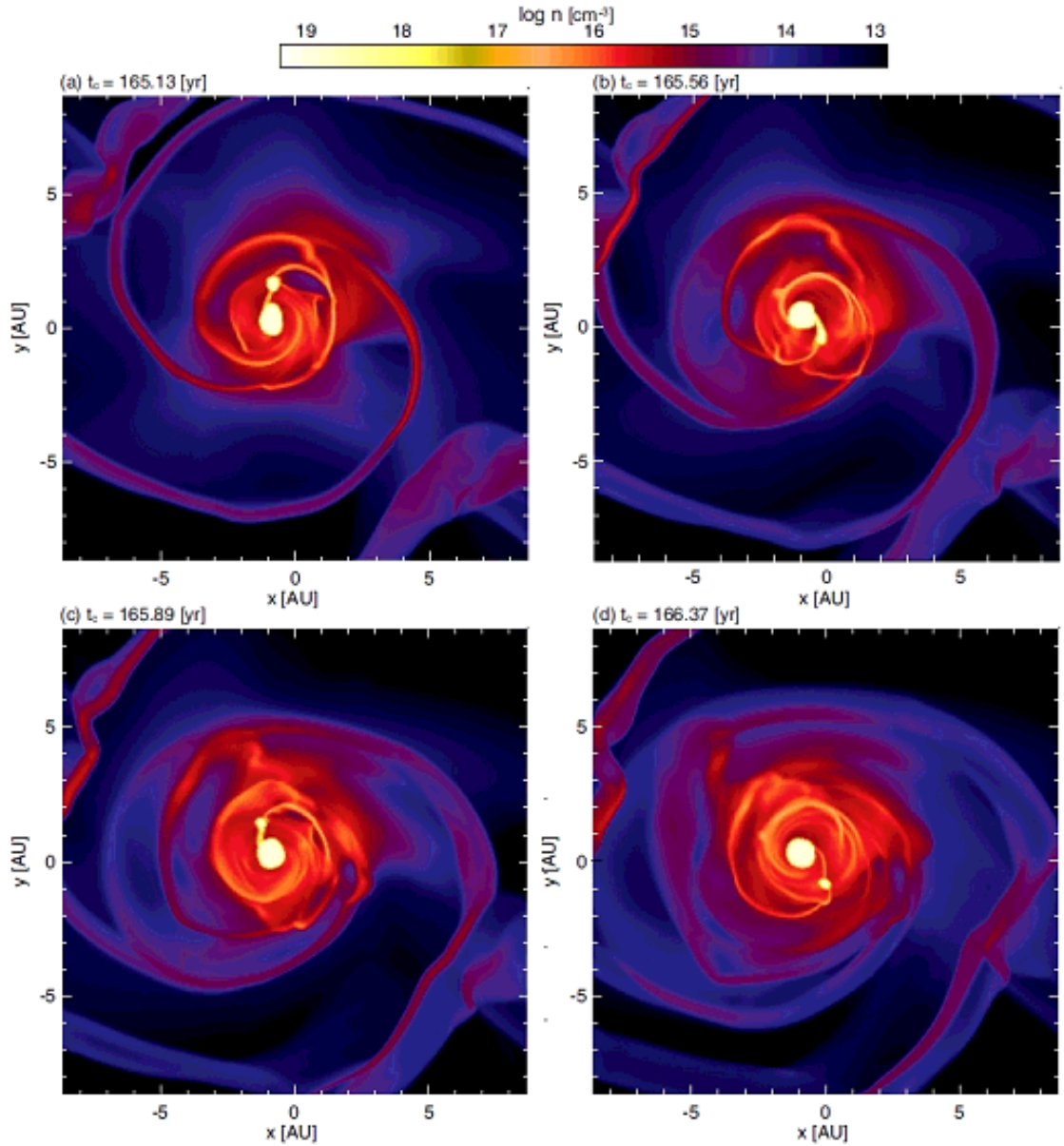}
\caption{
Density (colour) distributions around the centre of the cloud on the $z=0$ plane during $t_c=165.13$\,yr to $166.37$\,yr.
Primary star exchanges mass with small clump.
}
\label{fig:10}
\end{figure}

Figure~\ref{fig:9} plots the mass of primary and secondary stars against the elapsed time after primary star formation; the mass of the secondary star is plotted only when it exceeds $>0.01\msun$.
The primary star and other clumps have masses of $0.001-0.01\msun$ when they form.
The primary star acquires its mass from the circumstellar material.
The merger of small clumps also increases the primary stellar mass.
In the figure, a sudden drop in the mass of the primary star is caused by mass exchange between the primary star and other clumps.
When a clump approaches the primary star, it does not always merge into the primary star, as shown in Figure~\ref{fig:10}.
The clumps cannot merge with the primary star unless the orbital and spin angular momenta are effectively transferred outward.
Without effective angular momentum transfer, some of the gas in the primary star is stripped by the clump, and the mass of the primary star  decreases.
In contrast, the primary star sometime strips gas from the clump and increases its mass, as shown in Figure~\ref{fig:10}{\it b} and {\it c}.
As a result, the protostellar mass does not increase monotonically during the early phase of primordial star formation.
At the end of the calculation, the primary star has a mass of $8.8\msun$. 
In addition, there exist many less massive clumps having masses of $\lesssim 0.01\msun$. 
Such clumps are preferentially ejected from the centre of the cloud, as seen in Figure~\ref{fig:6}{\it g} and {\it h}.

In Figure~\ref{fig:9}, the mass decrease in the secondary star is caused by both gas stripping from the primary star and the falling onto the primary star.
During the early main accretion phase, fragmentation occurs only near the protostar ($r\lesssim 5$\,AU).
Thus, the clump orbits are comparable to the protostellar radius ($\sim 0.1-1$\,AU).
As a result, the secondary star (and other clumps) frequently interacts with the primary star. 
As the secondary star approaches the primary star, the gas in the secondary star is gradually stripped and finally falls onto the primary star.
Therefore, the secondary star does not grow sufficiently in this early phase.
However, with time, fragmentation occurs even in the region sufficiently far from the protostar, as shown in Figure~\ref{fig:6}. 
This is because the gas with a larger angular momentum later accretes onto the circumstellar region.
Although the mass of the secondary star is $0.8\msun$ at the end of the calculation, the secondary star and other clumps are expected to increase in mass with time because their orbits are far from the primary star and they rarely interact with other clumps and the primary star.
Note that in Figure~\ref{fig:9}, because a secondary star is ejected from the $l=20$ grid at $\tc\sim400$\,yr, we could not pursue the secondary star that is located in the region of $r\gg10$\,AU.

\begin{figure}
\includegraphics[width=150mm]{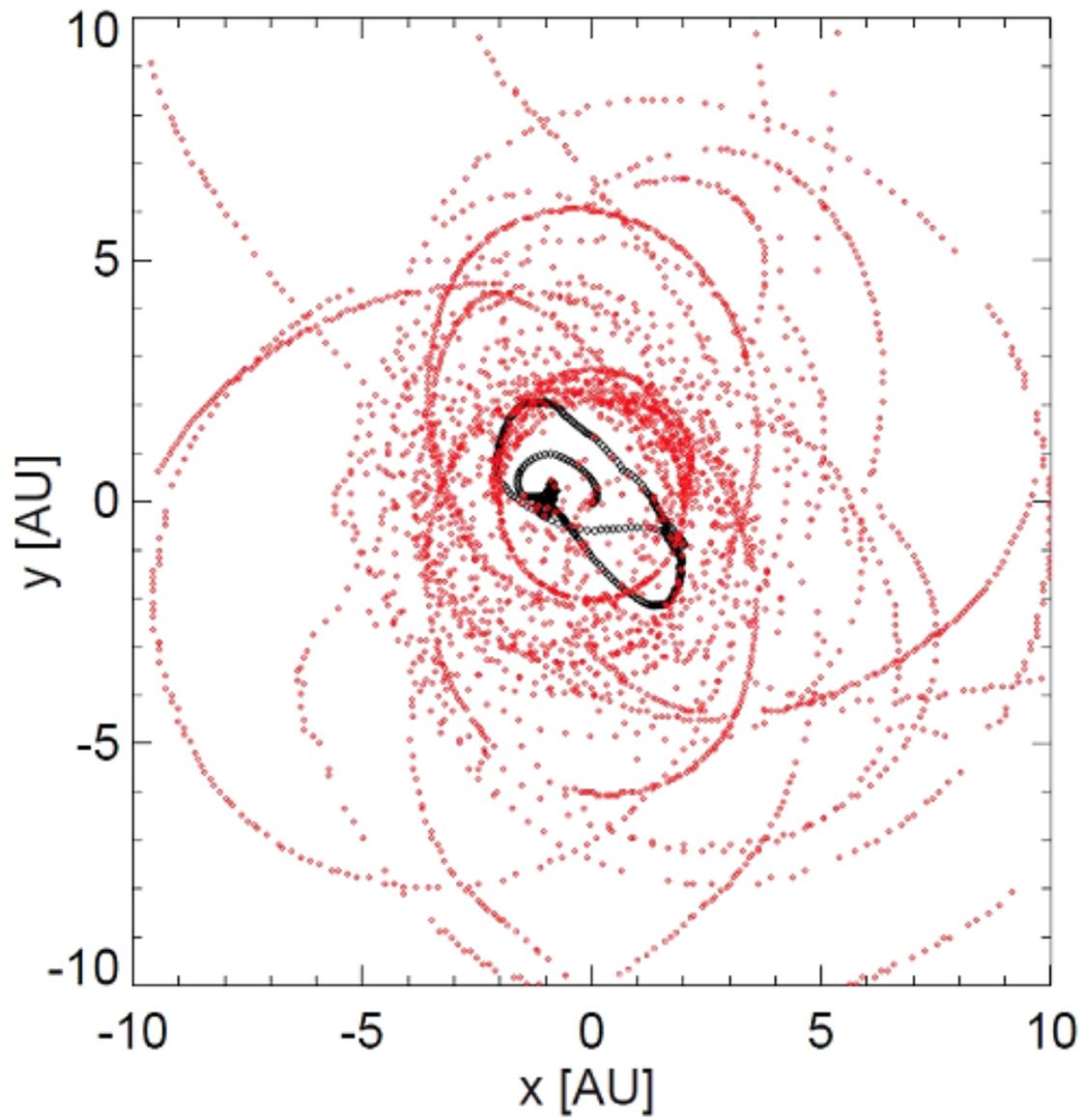}
\caption{
Trajectories for first formed protostar (or primary star; black dots) and later formed clumps (red dots) plotted on the equatorial plane.
}
\label{fig:11}
\end{figure}

The trajectories of clumps which have a mass of $>0.01\msun$ are plotted on the equatorial plane in Figure~\ref{fig:11}.
This figure indicates that many clumps orbit the primary stars, and some of them escaped from the central stellar system.
In total, nine clumps were ejected during the calculation.
Moreover, a secondary star with a mass of $0.8\msun$ was ejected from the centre of the cloud  with several less massive clumps at $t_c\simeq400$\,yr.
Thus, this system may evolve into a binary (or multiple) massive stellar system, in which some less massive stars orbit two massive stars.

By the end of the calculation, fragmentation continues to occurs, whereas the number of fragments around the primary star gradually decreases.
Since we did not calculate the cloud evolution for a long duration, we cannot judge whether the decrease in fragments is temporary.
As seen in Figure~\ref{fig:6}{\it i}, fragmentation rarely occurs, and a clear disk-like structure appears as the mass of the primary star increases.
A massive star can stabilize the disk against gravitational instability \citep{toomre64}.
Thus, a stable disk may appear in a further evolutionary stage of $\tc \gtrsim 1000$\,yr.

The cloud evolution after first protostar formation is very similar to that seen in \citet{greif12}, in which fragmentation frequently occurs around the primary star.
They calculated the evolution of a primordial cloud starting from the results of a cosmological simulation \citep{greif11} and resolved the protostar itself without any artifice using a highly developed numerical code that also included chemical reactions and radiative cooling. 
Although there are qualitative differences between our results and theirs, our results do not differ quantitatively from theirs. 
%%Thus, with our numerical code, we can also investigate the cloud evolution for magnetized primordial clouds.

\subsection{Evolution of Magnetized Primordial Cloud}
\label{sec:magnetized}
Fragmentation frequently occurs and many clumps appear in an unmagnetized primordial cloud, whereas only a single massive star appears in a (relatively strong) magnetized primordial cloud.
In this subsection, the cloud evolution of models with different magnetic field strengths (models 3-8) is described.
Among the models, the initial cloud has a magnetic field strength of  $10^{-5}$\,G $\le B_0 \le 10^{-10}$\,G, where $B_0$ is the magnetic field strength of the initial cloud that has a number density of $n=10^4\cm$.
Thus, the background magnetic field strength $B_{\rm amb}$ is expected to be weaker than $B_0$ because the background density is lower than $10^4\cm$, and the magnetic field is amplified as the density increases. 
In this subsection, we use $B_0$ as the initial field strength for convenience; we describe the corresponding background field strength in \S\ref{sec:summary}.

\begin{figure}
\includegraphics[width=150mm]{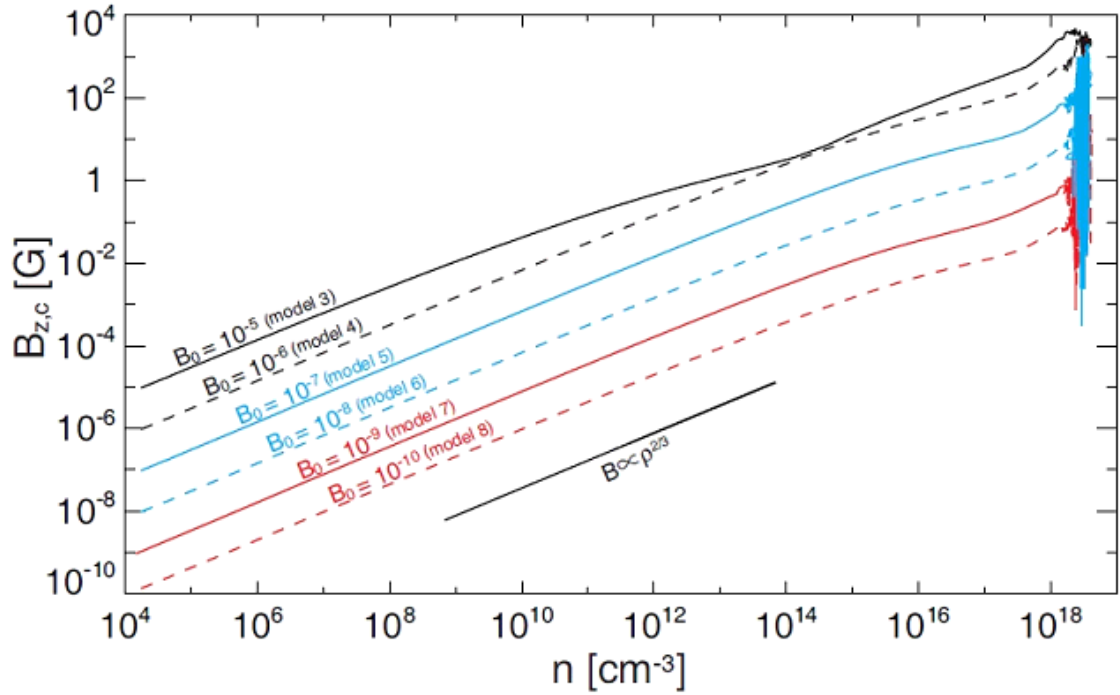}
\caption{
The $z$-components of the magnetic field strength at the centre of the cloud before protostar formation for models 3-8 against central number density.
The relation $B \propto \rho^{2/3}$ is also plotted.
}
\label{fig:12}
\end{figure}

The magnetic field strengths at the centre of the cloud for models 3-8 are plotted against the central number density in Figure~\ref{fig:12}.
The figure shows that, as the primordial cloud collapses, the magnetic field is amplified according to $B\propto \rho^{2/3}$, indicating that the cloud collapses almost spherically \citep{machida08c}.
Reflecting the initial difference in the magnetic field, the magnetic field strength of the protostar and its surrounding gas differ among the models.
At the protostar formation epoch, the magnetic field strength around the protostar is $B\sim10^3$\,G for model 3, but $B\sim0.1$\,G for model 8.
Thus, roughly speaking, the difference in the magnetic energy, which is proportional to $E_{\rm mag} \sim B^2$, is about 8 orders of magnitude among the models.

\begin{figure}
\includegraphics[width=150mm]{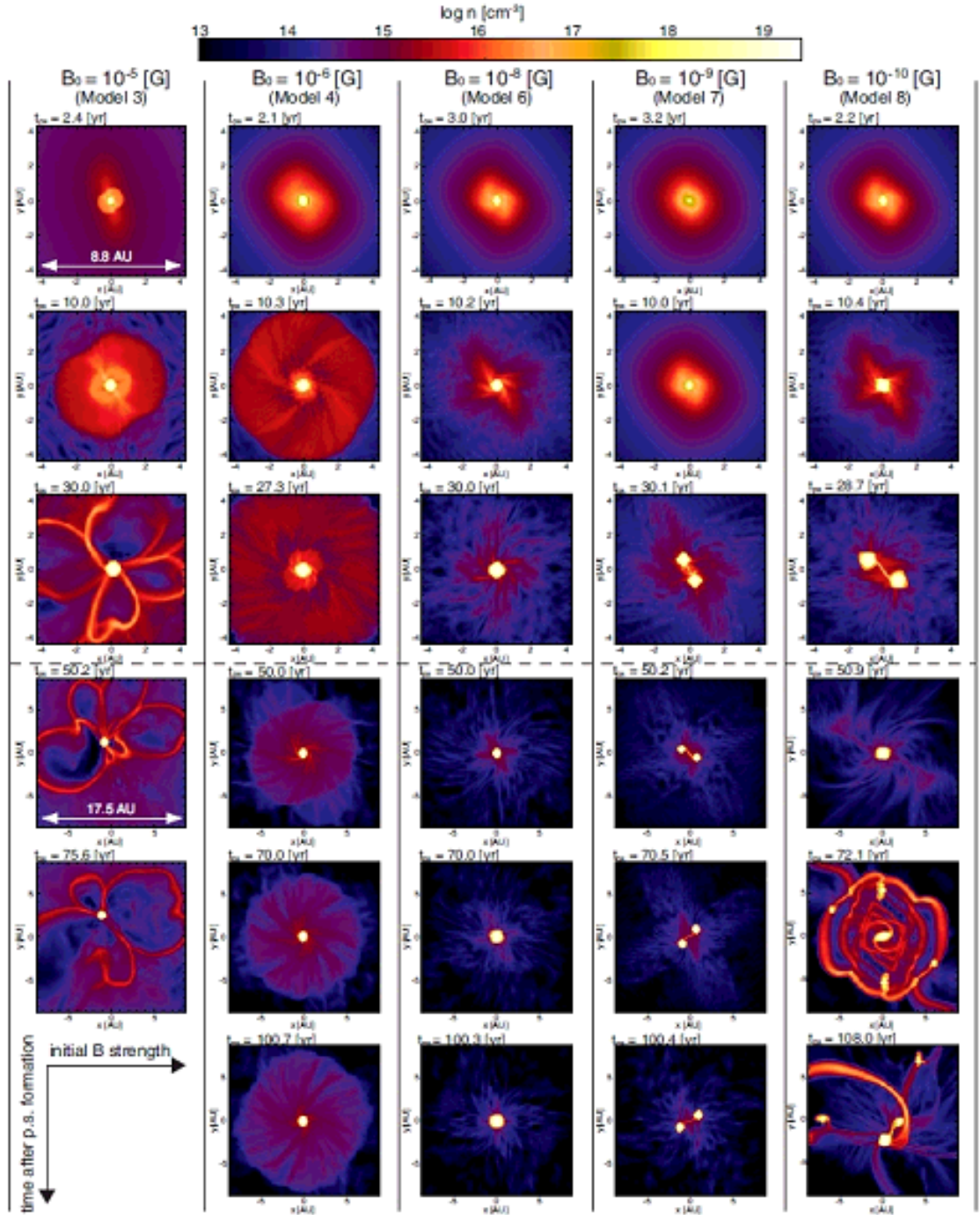}
\caption{
Time sequence images for models 3 (first column), 4 (second column), 6 (third column), 7 (fourth column) and 8 (fifth column).
In each panel, the density (colour) distribution  on the equatorial plane is plotted with a box size of $8.8$\,AU (first-third rows) and $17.6$\,AU (fourth-sixth rows).
Elapsed time is noted at the top  of each panel.
}
\label{fig:13}
\end{figure}

Figure~\ref{fig:13} shows time sequence images for the magnetized models (column from left to right: models 3, 4, 6, 7 and 8), in which the density distributions around the protostar at almost the same epochs of $t_c\simeq 2$, 10, 30, 50, 70 and 100\,yrs (each row) are plotted. 
No fragmentation occurs and a single protostar forms for models with an initially strong magnetic field (models 3, 4 and 6), whereas fragmentation occurs and a binary (model 7) or multiple (model 8) stellar system appears for models with an initially weak magnetic field.

Model 3 has the strongest initial magnetic field,  $B_0=10^{-5}$\,G, and its cloud evolution is plotted in the first column of Figure~\ref{fig:12}.
The figure shows that for model 3, a disk-like structure forms just after protostar formation ($t\lesssim10$\,yr).
We confirmed that the disk is supported not by rotation but by the magnetic field, and it has no azimuthal velocity component toward the protostar.
Thus, this disk corresponds to the so-called `pseudo-disk' \citep{galli93}.
For $t\gtrsim30$\,yr, filamentary structure develops around the protostar.
Without effective magnetic dissipation, the magnetic field accumulates in the protostar, and interchange instability can occur \citep{spruit90,spruit95,li96,stehle01,seifried11,krasnopolsky12,tomida13}.
Therefore, the magnetic field exudes from the protostar into the circumstellar region and forms a filamentary structure.
For model 3, we cannot confirm any sign of fragmentation by the end of the calculation.
In addition, we cannot calculate the cloud evolution beyond $t\sim80$\,yr after protostar formation for this model.
This is because the Alfv$\rm{\acute{e}}$n speed becomes quite high with a relatively strong magnetic field, and the time step becomes extremely small.
Thus, we cannot determine whether fragmentation occurs around the filaments in later evolutionary stages.
However, no fragmentation is expected because the angular momentum, which promotes fragmentation, is effectively transferred by magnetic effects.

For models 4, 5 and 6, only a single massive protostar appears without the filamentary structure. 
As seen in the second and third columns of Figure~\ref{fig:13}, no fragmentation occurs in spite of the appearance of a pseudo-disk around the protostar.
In addition, for these models, the angular momentum around the protostar is effectively transferred by magnetic braking.
Therefore, the infalling gas has very little angular momentum toward the protostar,  and it accretes directly onto the protostar.
Although the efficiency of the angular momentum transfer decreases as the magnetic field strength weakens, the initial magnetic field strength of $B_0\sim10^{-8}$\,G is sufficient to suppress fragmentation in the collapsing primordial cloud.

For model 7, fragmentation occurs and a binary system appears, as seen in the fourth column of Figure~\ref{fig:13}.
However,  no further fragmentation occurred by the end of the calculation.
Thus, even with an initially quite weak magnetic field strength of $B_0=10^{-9}$\,G, the magnetic field can transfer part of the excess angular momentum and influence the protostellar evolution.
After fragmentation, the binary separation gradually increases.
Thus,  part of the excess angular momentum is expected to contribute to increasing the orbital angular momentum of the binary system, which may suppress fragmentation and disk formation around binary systems.

For model 8, fragmentation occurs and many clumps appear, as observed in the unmagnetized model.
As seen in the fifth column of Figure~\ref{fig:13}, fragmentation occurs at $\tc \sim30$\,yr, and a binary stellar system appears.
Then, the binary stars merge to form a single star at $\tc\sim50$\,yr.
After the merger, a disk-like structure appears, and fragmentation frequently occurs.
As a result, many clumps form, as shown in the fifth row of the fifth column of Figure~\ref{fig:13}.
In the period of $\tc\gtrsim70$\,yr, the cloud evolution for model 8 is similar to that of the unmagnetized model (model 2), in which clump ejection and interaction between clumps often occur.  
As a result, the magnetic field rarely influences the protostellar evolution when the initial cloud has a magnetic field strength of  $B_0 \le 10^{-10}$\,G.

\begin{figure}
\includegraphics[width=150mm]{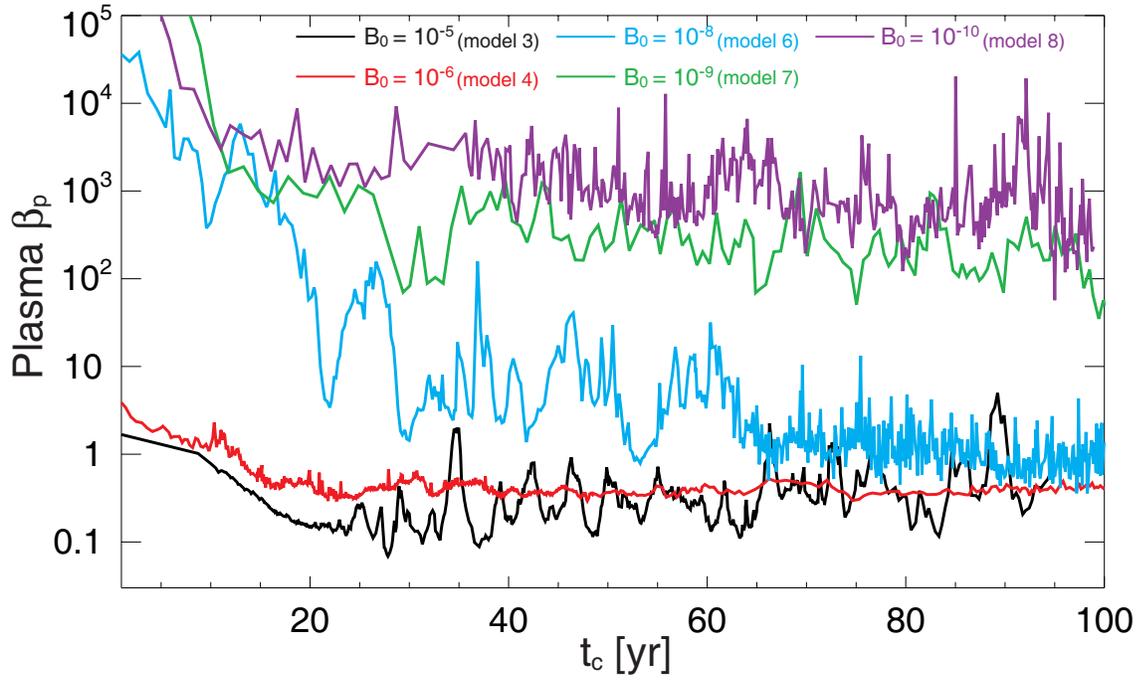}
\caption{
Plasma beta in the region of $r<5$\,AU for models 3, 4, 6, 7 and 8 against the elapsed time after protostar formation.
}
\label{fig:14}
\end{figure}

Figure~\ref{fig:14} shows the time evolution of the averaged plasma beta in the region of $r<5$\,AU, which is weighted by the density as
\begin{equation}
\beta_{p} = \dfrac{ \int (8 \pi P /B^2) \, \rho\, dv} {\int \rho \, dv},
\label{eq:beta}
\end{equation}
where $P$, $B$ and $\rho$ are the gas pressure, magnetic field strength and density, respectively, at each mesh point. 
Note that in equation~(\ref{eq:beta}), the plasma beta of the protostar (i.e. in the region of $n>10^{18}\cm$) is excluded in order to investigate the magnetic field strength of the circumstellar gas.
The figure indicates that models with an initially weak magnetic field tends to have a high plasma beta after protostar formation.
This is natural because the cloud with an initially weak field has a globally weak field  around the protostar (Fig.~\ref{fig:12}) before protostar formation, and the magnetic field near the protostar is amplified on the basis of this global field by the magneto-rotational instability (MRI, \citealt{balbus91}).
In other words, because the magnetic field strength amplification around the protostar depends on the global field \citep[e.g.][]{suzuki10}, a weaker magnetic field tends to be realized around protostars formed in clouds with a weaker magnetic field.
Note that shearing gas motion between the protostar and the circumstellar matter could contribute to the amplification of the magnetic field near the protostar.

As shown in Figure~\ref{fig:14}, models 3, 4 and 6 have a lower plasma beta in the range of $0.1\lesssim \beta_{\rm p} \lesssim 10$; they show no fragmentation (Fig.~\ref{fig:13}).
On the other hand, models 7 and 8 have $\beta_{\rm p} \gtrsim100$ and show fragmentation (Fig.~\ref{fig:13}).
This indicates that the magnetic field with $\beta_{\rm p} \lesssim 100$ can suppress fragmentation in primordial clouds.
In present-day star formation, the suppression of fragmentation by the magnetic field is reported in many studies \citep[e.g.,][]{machida05b,machida08a,price07,hennebelle08}.
This is because angular momentum (or rotation), which promotes fragmentation, is transferred by magnetic effects.

\begin{figure}
\includegraphics[width=150mm]{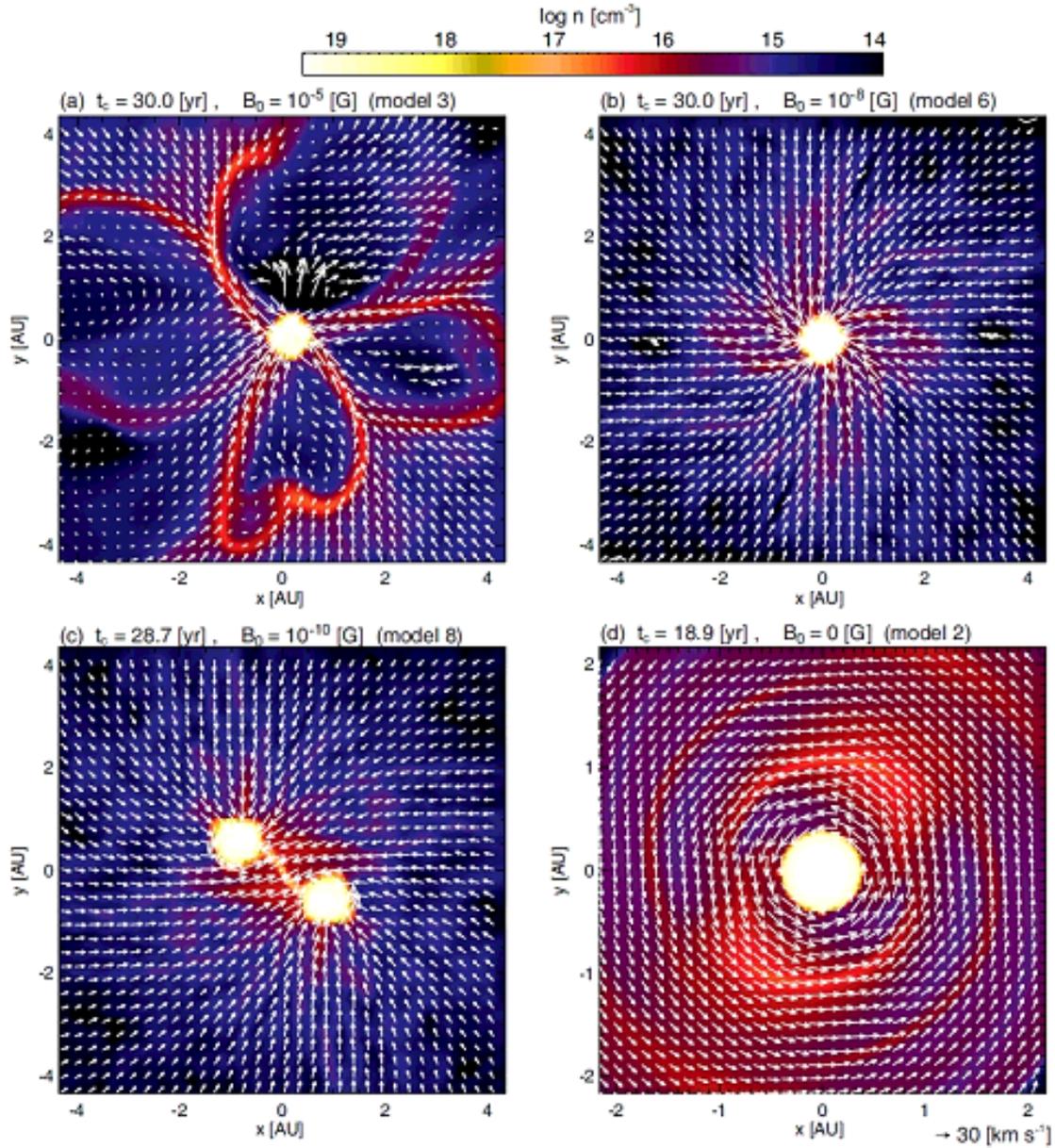}
\caption{
Density (colour) and velocity (arrows) distributions on the equatorial plane for models 2, 3, 6 and 8.
Elapsed time, initial magnetic field strength and model name are noted at the top of each panel.
}
\label{fig:15}
\end{figure}

To confirm the effects of the magnetic field around the protostar, the velocity and density distributions on the equatorial plane at $\tc \sim 20-30$\,yr for models 2, 3, 6 and 8 are plotted in Figure~\ref{fig:15}.  
Figure~\ref{fig:15}{\it a} indicates that, for model 3, the gas mainly flows into the protostar through the filament without rotation, whereas the gas from the low-density region flows out from the centre of the cloud. 
Note that the mass of outflowing gas on the equatorial plane is quite small.
The equatorial outflow can be confirmed in present-day star formation with a relatively strong magnetic field \citep{krasnopolsky10,krasnopolsky12}. 
In addition, in Figures~\ref{fig:15}{\it b} and {\it c}, we can confirm that the gas flows directly to the protostar without rotating for models 6 and 8, although a slight rotational motion is seen near the protostar in model 8.
In contrast, the rotational motion is outstanding for the unmagnetized model 2 (Fig.~\ref{fig:15}{\it d}).
In summary, around the protostar, the radial velocity component is strongly dominant over the azimuthal component in models with a magnetic field (models 3, 6, and 8), whereas the azimuthal component dominates in models without a magnetic field (model 2).
These models have the same initial rotational energy (Table~\ref{table:1}).
Thus, Figure~\ref{fig:15} clearly indicates that angular momentum is effectively transferred by the magnetic effect (or magnetic braking) in models with a magnetic field.
Because fragmentation occurs in a rotation-supported disk, it is natural that no fragmentation occurs in magnetized models, in which no rotation-supported disk forms.

In addition to magnetic braking, the protostellar jet also transfers the angular momentum around the protostar. 
After protostar formation, the jet appears in models 3, 4 and 5  among models 3--8(Table~\ref{table:1}). 
The jet is driven both by  magneto-centrifugally \citep{blandford82} and magnetic pressure gradient forces \citep{uchida85}.
Both a moderate magnetic field strength and moderate rotation are necessary to drive a jet in the collapsing gas cloud \citep{machida12}.
When the magnetic field is strong, the angular momentum is effectively transferred by magnetic braking and the reduced rotation is insufficient to drive a powerful jet.
A jet appears in model 3, which has the strongest magnetic field among the models, although it weakens and disappears in several tens of years after protostar formation.
In addition, models 6, 7 and 8  show no jet during the calculation because the weak magnetic field cannot drive it.
On the other hand, models 4 and 5  continue to show a powerful jet by the end of the calculation.

\begin{figure}
\includegraphics[width=150mm]{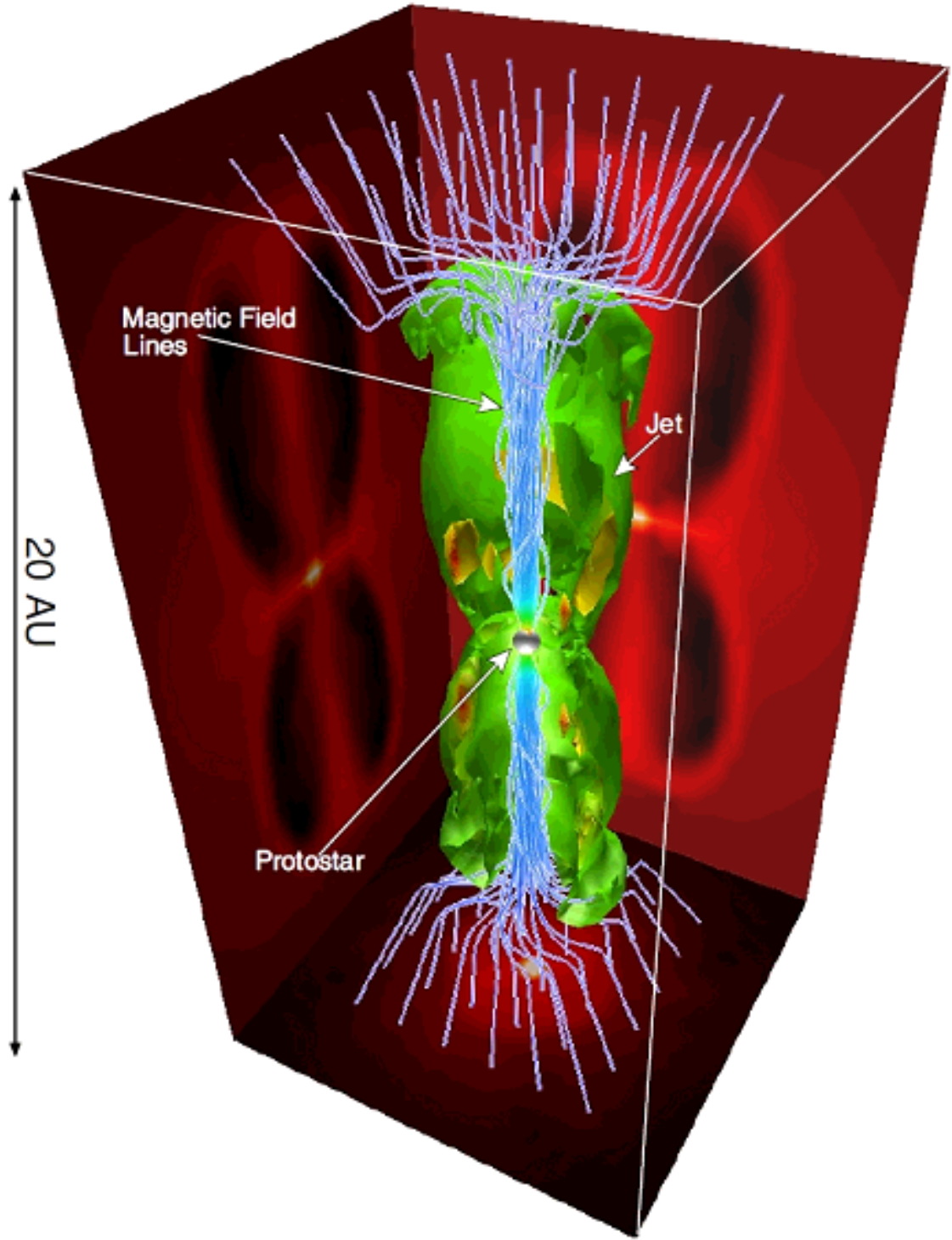}
\caption{
Three-dimensional structure of the jet for model 4.
The structures of the jet ($v_z > c_s$; green), protostar (grey) and magnetic field lines (blue and white streamlines) at $\tc=10$\,yr are plotted.
The density distributions on the $x=0$, $y=0$ and $z=0$ planes are projected on the wall surface. 
}
\label{fig:16}
\end{figure}

Figure~\ref{fig:16} plots the structure of the jet at $\tc=10$\,yr for model 4. 
For this model, a jet appears immediately after protostar formation.
The magnetic field lines are strongly twisted by the rotation of the protostar, and drive a well-collimated jet.
The jet has a maximum velocity of $79\km$, which roughly corresponds to the Kepler velocity when the protostar has a mass of $\sim2\msun$ and a radius of $60\,\rsun$.
The jet has a mass of $\sim0.14\msun$ at the end of the calculation ($\tc\simeq100$\,yr); thus, the mass ejection rate is estimated as $\sim0.014\mdot$.
About 10--20\% of the accreting matter is ejected by the protostellar jet.

\section{Discussion}
\subsection{Effect of Initial Rotational Energy}
In \S\ref{sec:results}, we described the cloud evolution when the initial cloud has a constant rotational energy of $\beta_0=10^{-4}$.
%% that nearly corresponds to the minimum of molecular cloud core in the solar neighborhood \citep{caselli02}.
The rotation promotes fragmentation during the gas collapsing phase in primordial clouds \citep{machida08,machida08c}.
Thus, fragmentation is thought to occur more frequently even during the gas accretion phase with a larger rotational energy when the primordial cloud is unmagnetized \citep{greif12}.
In magnetized clouds, fragmentation is suppressed by  magnetic effects.
At $\beta_0=10^{-4}$, no fragmentation occurs when the initial cloud has a magnetic field of $B_0 \ge 10^{-8}$\,G.
However, the conditions for fragmentation are also expected to depend on the initial rotational energy $\beta_0$.

\begin{figure}
\includegraphics[width=150mm]{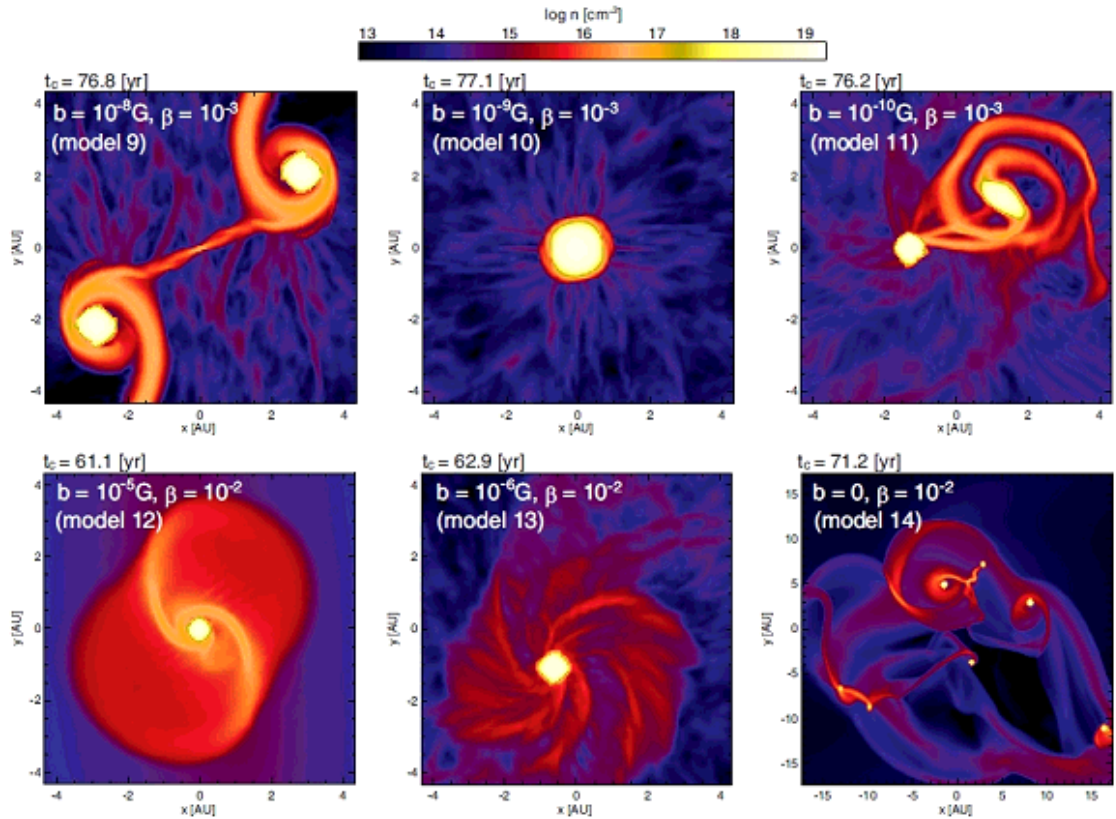}
\caption{
Density (colour) distribution on equatorial plane for models 9-14.
Elapsed time after protostar formation, parameters ($b$ and $\beta$) and model name are given in each panel.
}
\label{fig:17}
\end{figure}

To investigate the effect of the initial rotation on fragmentation, we also calculated the cloud evolution for models with the rotational energy of $\beta_0=10^{-2}$ and $10^{-3}$ (models 9--14).
Figure~\ref{fig:17} shows the density distribution around the protostar at $\tc \sim 60-80$\,yr for these models.
Models 9--11 have the same strength of the initial magnetic field as in models 6--8, but their initial rotational energy is 10 times larger, as shown in Table~\ref{table:1}.
With $B_0=10^{-8}$\,G, fragmentation does not occur for model 6 ($\beta_0=10^{-4}$), whereas it  does occurs for model 9 ($\beta_0=10^{-3}$).
Thus, it seems that fragmentation can occurs even in a strongly magnetized cloud when the initial cloud has  larger rotation energy.

In contrast to models with $B_0=10^{-8}$\,G, when $B_0=10^{-9}$\,G,  fragmentation did not occur in models with both $\beta_0=10^{-4}$ (model 7) and $10^{-3}$ (model 10).
As seen in Figures~\ref{fig:13} and \ref{fig:17}, only a single protostar exists by the end of the calculation  in both models.
When the magnetic field is {\it somewhat} weak at the protostar formation epoch, the angular momentum is not so effectively transferred outward and a rotation disk (temporarily) forms.
The magnetic field is effectively amplified in the rotating disk by the MRI and shearing gas motion, and the amplified magnetic field further transfers the angular momentum of the disk.
Thus, the amplified field suppresses fragmentation.
Therefore, after protostar formation (or rotating disk formation), the relation between the fragmentation condition and the magnetic field and rotation is not so simple.
For this reason, we cannot guarantee that a rapid rotation always promotes fragmentation in magnetized clouds during the gas accretion phase.
When the initial cloud has a magnetic field of $B_0=10^{-10}$\,G, fragmentation occurs in models with both $\beta_0=10^{-3}$ (model 8) and $10^{-4}$ (model 11).
Thus, when magnetic field strength is sufficiently weak, fragmentation seems to occur regardless of the initial rotational energy.

In the cloud with larger rotational energy of $\beta=10^{-2}$ and strong magnetic field of $B_0=10^{-5}$\,G (models 12 and 13), only a single protostar exists at $\tc \simeq 60$\,yr.
No fragmentation occurs for model 12, whereas fragmentation occurs just after the first protostar formation for model 13.
In model 13, although four protostars appear after fragmentation, protostars are merged into a single protostar at $\tc\sim50$\,yr as seen in Figure~\ref{fig:17}.
Although we adopted a stronger magnetic field for models 12 and 13 because we have to limit the number of models (\S\ref{sec:initial}), the magnetic field seems to induce a single Population III star formation even in the clouds with a larger rotational energy. 
On the other hand, as shown in Figure~\ref{fig:17}, fragmentation occurs in a large scale for unmagnetized model 14 that has the same initial rotational energy as in models 12 and 13.
In summary, larger rotational energy tends to promote fragmentation (in a large scale), whereas larger magnetic energy tends to suppress it. 
However, because the magnetic field is amplified by the rotation and transfers angular momentum, we cannot simply determine the fragmentation condition in the gas accretion phase.
Further calculations are necessary to clarify it.

\subsection{Effect of Initial Magnetic Configuration}
\label{sec:mag-config}
In this study, we assumed large-scale ordered magnetic fields for magnetized models.
We can understand the configuration and strength of magnetic fields in nearby star forming regions by  observations of the polarization pattern and Zeeman splitting.
The observation by \citet{li09} showed that  small-scale (or cloud core scale) magnetic fields are correlated or aligned with  large scale fields \citep[see also][]{tamura87,tamura89,girart06,girart09}. 
Thus, we usually adopted large-scale ordered magnetic fields to investigate the formation of present-day stars \citep[e.g.][]{machida06, commercon10, seifried11, tomida13}.
On the other hand, the configuration and strength of magnetic fields in the early universe is controversial, because we cannot directly observe them.
However, there are some theoretical implications.
Recent studies claimed that the accretion shock onto the first star-forming halo can create turbulence which generates weak seed magnetic fields being amplified by the small-scale dynamo \citep[e.g.,][]{schleicher10,sur10,federrath11}. 
Thus, to investigate the star formation in the early universe, we may have to adopt small-scale disordered fields in turbulent clouds according to the latest theoretical prediction.

When the disordered fields are adopted, local magnetic field lines are not aligned with the local rotation axis. 
In such a case, the cloud evolution somewhat differs from that in the aligned case \citep{matsu04,machida06}.
\cite{joos12} calculated the disk formation when the global magnetic fields are misaligned to the global rotation axis and pointed out that misaligned fields weaken the efficiency of the magnetic braking and promote disk formation \citep[see also][]{hennebelle09,joos13}.
\citet{commercon11} calculated the evolution of turbulent cloud with global ordered fields and showed  the suppression of fragmentation during the early phase of the star formation. 
\citet{seifried12} calculated the cloud evolution with almost the same setting as in \citet{commercon11} and claimed that the magnetic braking efficiency is reduced due to the absence of a coherent rotation structure around the center of the collapsing cloud, which  is caused by turbulence.
In addition, turbulent reconnection may also weaken the magnetic effects on the disk formation \citep{Santos-Lima12}.
Thus, it is expected that disordered small scale fields alleviate the effect of the magnetic field or efficiency of magnetic braking. 
In this study, we adopted ordered global fields as a first attempt at the problem, and will postpone the investigation of disordered fields to a later paper. 
%%In this study, although we adopted ordered global fields as the first step to study the 
%%effect of the magnetic field on Population III star formation, we should investigate it with 
%%disordered fields and turbulence in  future.

\subsection{Protostellar Mass Evolution}
\begin{figure}
\includegraphics[width=150mm]{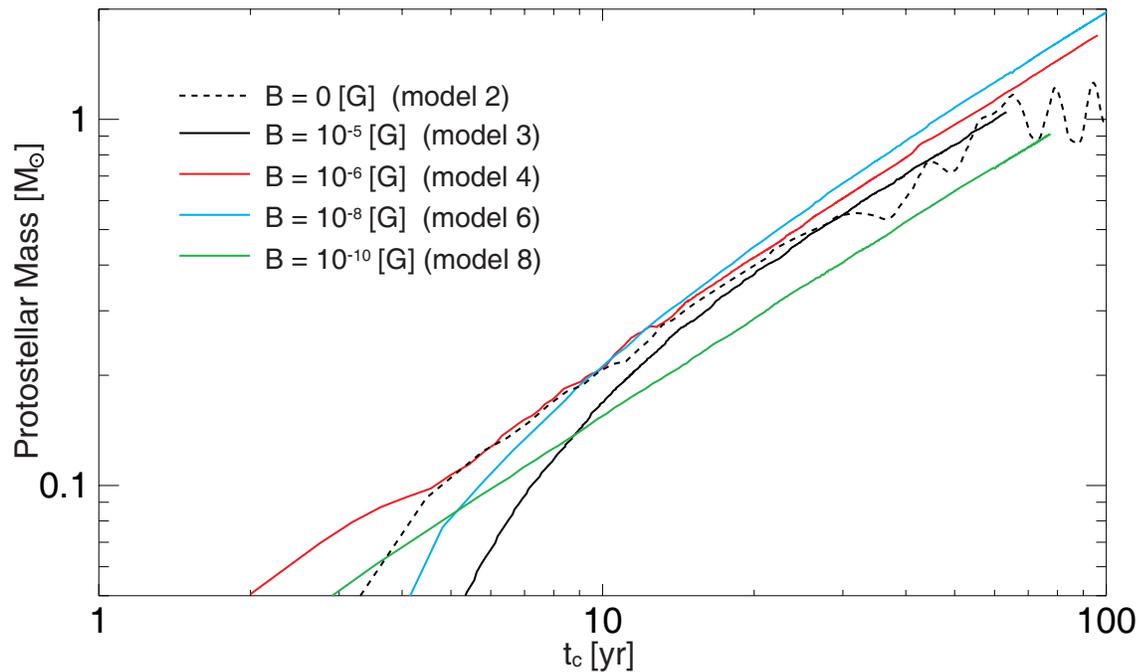}
\caption{
Primary protostellar mass for models 2, 3, 4, 6 and 8 against the elapsed times after protostar formation.
}
\label{fig:18}
\end{figure}

The magnetic field affects the environment around protostar because it suppresses disk formation and fragmentation and drives the protostellar jet.
To investigate the effect of the magnetic field on protostellar growth, the mass of the primary star for models 2, 3, 4, 6 and 8 is plotted against the elapsed time after primary star formation in Figure~\ref{fig:18}.
The primary star corresponds to a most massive clump that has a maximum density of $n>10^{18}\cm$ (\S\ref{sec:unmanetized}), and  the mass of the primary star is calculated integrating the gas in the region of $n>10^{18}\cm$.
%%and the mass of the primary star is calculated integrating the gas in the region of $n>10^{18}\cm$ inside the clump.
The figure shows that the difference in the primary stellar mass among the models is within a factor of $\sim 2$.
This indicates that the magnetic field has little effect on the mass accretion rate onto the primary star.
In present-day star formation,  mass accretes onto the protostar through the circumstellar disk.
Thus, the mass accretion rate is strongly related to the condition (or viscosity) of the circumstellar disk.
On the other hand, no clear circumstellar disk appears in a primordial cloud.
When the initial cloud is unmagnetized, fragmentation frequently occurs and fragments break up the disk.
In addition, when the initial cloud is magnetized, the magnetic field effectively transfers the angular momentum, and no disk forms.
As a result,  gas accretes directly onto the protostar from the infalling envelope.
The density profile of the collapsing prestellar cloud, which roughly corresponds to the infalling envelope after protostar formation, is described well with a simple self-similar solution even with rotation and a magnetic field as described in \S\ref{sec:initial}.
Thus, it is natural that the models exhibit almost the same  mass accretion rate.
In an unmagnetized cloud, the protostar also acquires mass from clump mergers and loses mass by mass exchange, as described in \S\ref{sec:unmanetized}.
However, Figure~\ref{fig:18} indicates that mergers and mass exchange do not contribute greatly to the mass evolution of the primary star.

As described in \S\ref{sec:results}, many clumps appear in the unmagnetized cloud, whereas only a single protostar appears in the magnetized cloud.
However, the primary star has almost the same mass in any model.  
In addition, the mass accretion rate of the primary star for all models is as high as $\dot{M}_* \sim 10^{-3}-10^{-2}\mdot$ during the calculations.
Moreover, the primordial cloud has sufficient mass to supply the protostar in further evolutionary stages.
Although mass accretion finally stops owing to protostellar feedback \citep{hosokawa11a,hosokawa12}, a massive protostar of $\gg 10\,\msun$ is expected to appear in both unmagnetized and magnetized primordial clouds.

In addition, when fragmentation frequently occurs in the disk, ``fragmentation induced starvation''  limits the protostellar mass evolution \citep{peters10b}. 
\citet{peters10a} calculated the massive star formation in a massive cloud. 
They  pointed out that the accretion onto the central star is shut off by disk fragmentation and the formation of lower-mass companions that can intercept the inward mass accretion. 
In our calculation, the protostellar mass continues to increase for at least $\sim$100\,yr after the first protostar formation.
However, in  a further evolutionary stage, the growth rate of the protostellar mass may decrease by this mechanism.
To determine the final stellar mass, we need to calculate the cloud evolution for a longer time.

\subsection{Comparison with Present-day Star Formation}
\subsubsection{Fragmentation Scale}
In this study, we calculated the evolution of primordial clouds and investigated the star formation process in the main accretion stage.
In the calculations, fragmentation frequently occurs when the primordial cloud is unmagnetized or very weakly magnetized, whereas no fragmentation occurs when it is (strongly) magnetized.
Consequently, many stars and substellar companions appear in the former case, whereas only a single massive star appears in the latter case.
These features of star formation in primordial clouds differ in several aspects from those in  present-day clouds.
It is useful to clarify the difference in the star formation process between primordial and present-day clouds.

In present-day clouds, gas evolves almost isothermally for $n\lesssim 10^{11}\cm$.
After the gas becomes optically thick toward  dust thermal emission at $n \sim 10^{11}\cm$, it evolves adiabatically for $n \gtrsim 10^{11}\cm$ (see Fig.~\ref{fig:1}), and the  first adiabatic core forms.
Next, the density increases very slowly because a quasi-equilibrium state is realized.
Thus, the first adiabatic core has enough time to develop the perturbations which cause fragmentation.
Note that, without turbulence, because the gas collapses very rapidly and there is not enough time to develop perturbations before the first adiabatic core forms,  fragmentation does not occur unless the molecular cloud core is extremely distorted \citep{tsuribe99}.
Note also that, with turbulence, fragmentation can occur even in the isothermal collapsing phase \citep{goodwin04a, goodwin04b,walch12,seifried12}. 
As a result, without turbulence, fragmentation can occur on the scale of a first core which is 10-100\,AU in size.
Then, the protostar forms $\sim 10^2-10^4$\,yr after first core formation.
The first core (or the remnant of the first core) remains after protostar formation and becomes the circumstellar disk \citep{bate98,bate11,machida10a}.
Thus, the circumstellar disk has a size of $\sim 10-100$\,AU at the protostar formation epoch.
During the main accretion phase, fragmentation occurs in the outer region of the disk \citep[e.g.][]{stamatellos07}.
In addition, fragmentation rarely occurs at $r <1$\,AU before and after protostar formation.
In summary, fragmentation occurs at $r\gg1$\,AU in present-day star formation.
Thus, we may introduce a sink with an accretion radius of $r\lesssim 1$\,AU.

On the other hand, the gas pressure (or temperature) gradually increases with $P \propto \rho^{\gamma}$ ($T\propto \rho^{\gamma-1}$) and $\gamma \sim 1.1$ in primordial clouds (Fig.~\ref{fig:1}).
Unlike the case in present-day clouds, in the primordial clouds, no adiabatic core forms before protostar formation because the adiabatic index $\gamma$ never exceeds $\gamma>4/3$  until the protostar formation.
However, fragmentation can occur in the collapsing primordial cloud because there is no analytical solution in a rotating cloud with a polytropic index of $\gamma=1.1$ \citep{saigo00}.
Note that there exists a self-similar solution with $\gamma=1$, in which no fragmentation occurs because the gas collapses according to the solution.
Although fragmentation manages to occur before protostar formation \citep{turk09}, it easily occurs after protostar formation because the gas collapse stops and the perturbations necessary for fragmentation can develop.
Because the dynamical timescale (free-fall timescale or rotation timescale) is short near the protostar, fragmentation tends to occurs near the protostar in the early phase of star formation.
In addition, gas with a lower angular momentum, which has a smaller centrifugal radius \citep{cassen81}, first falls near the protostar and contributes to fragmentation.
In reality, as seen in Figure~\ref{fig:6}, fragmentation occurs only near the primary protostar, at least during the early gas accretion phase.
 Note that fragmentation may not occurs in the very proximity of the protostar, because the accretion luminosity heating, which is not considered in this study, can suppress fragmentation in such a region after the protostar becomes sufficiently massive \citep{clark11b,smith11,greif11}.

In summary, for the primordial case, only the (primary) protostar can give a typical scale for fragmentation because there is no typical scale in the collapsing cloud with an adiabatic index of $\gamma=1.1$.
If a sink cell is introduced in the primordial cloud, the sink radius (or accretion radius) becomes the typical scale for fragmentation, and fragmentation occurs artificially with the sink radius, as described in the Appendix.
Thus, we cannot use a sink for the primordial star formation process.
On the other hand, we may use it for present-day star formation because the typical scale appears before protostar formation with the first adiabatic core.

\citet{greif12} pointed out that, in primordial clouds, fragmentation does not occur in the range of $n>10^{17}\cm$, with which we can estimate the typical fragmentation scale (or Jeans length) of $\lambda_{J}\simeq0.2(T/2000\,K)^{1/2} (n/10^{17}\cm)^{1/2}$\,AU. 
Thus, we may investigate fragmentation with the sink accretion radius of $r_{\rm acc}<0.2$\,AU. 
Note that we need to be slightly careful when using a sink (particle) implementation, since the protostellar radius changes from 0.1AU to 1\,AU as the protostar evolves.

\subsubsection{Mass Accretion Rate and Fragmentation}
The mass accretion rate onto the protostellar system (circumstellar disk and protostar) differs considerably between primordial and present-day clouds: the mass accretion rate is $\dot{M}\sim0.01\mdot$ for primordial clouds, whereas it is $\dot{M}\sim10^{-6}\mdot$ for present-day clouds.
Thus, in the primordial cloud, the disk surface density fluctuates over a short time period.
The surface density is directly related to the gravitational instability and fragmentation.
The gravitational instability of the disk is described using Toomre's Q parameter, which is defined as 
\begin{equation}
Q = \dfrac{c_s\, \kappa}{\pi \, G \, \Sigma},
\label{eq:toomre}
\end{equation}
where $\kappa$ and $\Sigma$ are the epicyclic frequency and surface density, respectively.
The high mass accretion rate rapidly increases the disk surface density, and the denominator of equation~(\ref{eq:toomre}) increases.
Then, $Q$ becomes as small as $Q<1$, and gravitational instability or fragmentation occurs.
Assuming a disk mass of $M_{\rm d}\sim0.01\msun$ and a disk mass accretion rate of $\dot{M}_{\rm d} \sim0.01\mdot$ for the primordial case and $\dot{M}_{\rm d} \sim 10^{-6}\mdot$ for the present-day case, the disk growth timescale is 
\begin{eqnarray}
t_{\rm grow} \equiv \left( \dfrac{M_{\rm d}}{\dot{M}_{\rm d}} \right) \sim 
\left\{
\begin{array}{ll} 
1  \ {\rm yr} \ \ \ \ {\rm for \ \ primordial \ clouds}\\
10^4  \ {\rm yr}  \ \ \ {\rm for \ \ present\mathchar`-day \ clouds}, \\
\end{array}
\right.
\end{eqnarray}
In addition, the Keplerian timescale is described as
\begin{equation}
t_{\rm Kep} \sim \dfrac{2\pi}{\Omega} = \sqrt{\dfrac{4\pi^2 r^3}{G\, M}  }.
\label{eq:kepler}
\end{equation}
As seen in Figure~\ref{fig:6}, fragmentation occurs at $r\sim1-5$\,AU in primordial clouds.
When $r=5$\,AU (or 1\,AU) and $M=1\msun$ are substituted into equation~(\ref{eq:kepler}), the Keplerian timescale is $t=11.2$\,yr (or 1\,yr).
For present-day cloud, the Keplerian timescale $t_{\rm Kep}\sim1-10$\,yr at 1-5\,AU is much shorter than the disk growth timescale $t_{\rm grow}\sim10^4$\,yr.
In this case, the disk may adjust itself by developing a spiral structure which can transfer its angular momentum and redistribute surface density.
When the spiral structure cannot transfer sufficient angular momentum, fragmentation may occur.
In addition, the cooling timescale should be taken into account \citep{gammie01} because the denominator (or sound speed) in equation~(\ref{eq:toomre}) also changes on a timescale of $\sim10^4$\,yr \citep[$\sim t_{\rm cool}$;][]{inutsuka10}.
On the other hand, for primordial clouds, the Keplerian timescale $t_{\rm Kep} \sim 1-10$\,yr is longer than or comparable to the disk growth timescale, $t_{\rm grow}\sim1$\,yr. 
Hence, fragmentation is unavoidable because the disk grows before it can redistribute its angular momentum and surface density. 
Note that the cooling timescale is much longer than either the Keplerian or disk growth timescale.
As a result, at a higher mass accretion rate, fragmentation tends to occur in primordial clouds.

\subsubsection{Magnetic Braking Catastrophe} 
A magnetic field impedes disk formation.
The disk is supported by rotation, and its angular momentum is transferred by magnetic braking. 
Thus, when the magnetic field effectively transfers the angular momentum outward, no disk forms around the protostar. 
In present-day star formation, two processes (magnetic dissipation and a limited mass reservoir) make disk formation possible, overcoming the magnetic braking catastrophe \citep{mellon08,mellon09,li11,machida11c}.

For present-day collapsing clouds, the magnetic field in the high-density gas region effectively dissipates by Ohmic dissipation with a much lower ionization degree.
The magnetic dissipation weakens the efficiency of angular momentum transfer by magnetic braking. 
However, it requires appreciable time to dissipate the magnetic field. 
The first core (the first core remnant), which  exists for $10^3-10^4$\,yr even after protostar formation, plays a crucial role for the Ohmic dissipation \citep{machida11c}. 
Because the infalling gas stays in the first core (remnant) for a long time, the magnetic field in the first core can dissipate, and a rotation-supported disk forms.
In addition, as the infalling envelope dissipates,  magnetic braking becomes ineffective, and the disk grows. 
Note that for magnetic braking to occur, the infalling envelope must have sufficient mass because angular momentum is transferred into the infalling envelope, which can brake the disk.

On the other hand, as shown in Figure~\ref{fig:2}, Ohmic dissipation is not so effective in primordial clouds. 
If the gas can stay in the disk, Ohmic dissipation may decrease the magnetic field strength.
However, the disk (or first core) does not form in primordial clouds.
Thus, gas cannot stay in the circumstellar region without dissipation of the magnetic field, and angular momentum is effectively transferred by magnetic braking.
In addition, because the primordial cloud has a sufficient mass of $\sim10^3-10^4\msun$, the infalling envelope is not depleted.
Such a massive infalling envelope can continue to brake the disk (or transfer angular momentum outward) even long after protostar formation.
Note that, however, as the protostar becomes massive enough, the infalling envelope may be dissipated by protostellar feedback.
In summary, the magnetic braking catastrophe in primordial clouds is more serious than that in present-day clouds.
Thus, a rotation-supported disk cannot form when a primordial cloud is magnetized.

\section{Summary}
\label{sec:summary}
In this study, we investigated the accretion phase of Population III star formation in magnetized and unmagnetized primordial clouds.
First, to make a long-term calculation after protostar formation possible, we constructed a protostellar model by adjusting the equation of state, which relates the protostellar mass  to the protostellar radius.
Then, we calculated the evolution of primordial magnetized and unmagnetized clouds for several hundreds of years after first protostar formation.
Our calculations yielded the following results.

\begin{itemize}
\item {\bf Multiple Fragments and Stellar Clusters in Unmagnetized Clouds}\\
In unmagnetized clouds, fragmentation frequently occurs after the formation of the first protostar, which becomes the primary (or most massive) star.
A disk-like structure appears just after first protostar formation, whereas it fragments into several clumps.
Thus, although the infalling gas tends to form a disk-like structure around the primary star,  prompt fragmentation breaks up the disk.
The mass accretion rate in primordial clouds is as high as $\dot{M}\sim 10^{-2} \msun$\,yr$^{-1}$, which is about 3 -- 4 orders of magnitude higher than that in present-day clouds.
With this high mass accretion rate, the disk rapidly becomes gravitationally unstable, and fragmentation occurs before the development of sufficient non-axisymmetric perturbation, which can redistribute the angular momentum in the disk and suppress fragmentation. 
As a result, many clumps appear in the collapsing primordial cloud.
Although some clumps fall onto the primary star, many of them survive.
Thus, it is thought that a multiple stellar system, or stellar cluster, appears in unmagnetized  primordial clouds.
In addition, some of the clumps are ejected from the centre of the cloud with masses ranging from $\sim 0.01\msun $ to $0.1\msun$. 
These clumps are expected to become metal-free low mass stars as already shown in previous studies \citep{clark08,smith11,clark11b,greif12}. 
\\

\item {\bf Effects of Magnetic Field}\\
When the primordial cloud is magnetized, no fragmentation occurs and a single massive star forms.
The magnetic field is amplified in proportion to $B \propto \rho^{2/3}$ in a spherically collapsing cloud. 
The density contrast between the first collapsed object or the initial cloud ($ n \sim 10^3-10^4 \cm$) and the protostar ($n \gtrsim 10^{18}\cm$) is about $10^{15}$  orders of magnitude. 
Thus, the magnetic field around the protostar at its formation is about $10^{10}$ times stronger than that of the first collapsed object.
In addition, after protostar formation, the magnetic field can be amplified by both MRI  and shearing motion between the protostar and infalling gas.
Moreover, the magnetic field in primordial clouds does not dissipate significantly because of the high gas temperature and absence of dust grains.
Therefore, the circumstellar gas near the protostar has a non-negligible magnetic field strength even when the magnetic field is quite weak in a prestellar cloud.
Around the protostar, the magnetic field can effectively transfer angular momentum by both magnetic braking and protostellar jets. 
Thus, gas can fall directly onto the protostar without forming a rotation-supported disk.
As a result, only a single massive star appears in a magnetized primordial cloud. 
\\

\item {\bf Single Star or Stellar Cluster}\\
Although the magnetic field suppresses disk formation and subsequent fragmentation, an extremely weak field cannot affect the dynamical evolution of a primordial cloud. 
To quantitatively estimate the effects of a magnetic field, we calculated the evolution of primordial clouds with different parameters of the initial magnetic field.
Our calculation showed that the magnetic field can affect the formation and evolution of Population III stars when the plasma $\beta$ around  the protostar is $\beta\lesssim 10-100$, which is realized when the initial cloud, which has a number density of $10^4\cm$, has a magnetic field strength of 
$B_0 \gtrsim 10^{-10}$\,G.
We can also estimate the magnetic field strength of the ambient medium using the relation $B \propto \rho^{2/3}$ (Table~\ref{table:1}).
%% which corresponds to $B \lesssim 2 \times 10^{-13}(n/1 \cm)^{-2/3} $G with the relation .
Assuming a hydrogen number density of $n=1\cm$, the Population III star formation in magnetized clouds can be classified as
\[ 
\left\{
\begin{array}{l}
  {\rm Single \ Massive \ Star:} \   B_{\rm amb}  \, \gtrsim \, 10^{-12}\, (n/1\cm)^{-2/3} \ {\rm G}       , \\
  {\rm Binary \ System:} \   10^{-12}\, (n/1 \cm)^{-2/3} \ {\rm G} \, \lesssim \, B_{\rm amb} \,  \lesssim \, 10^{-13}\, (n/1\cm)^{-2/3} \ {\rm G} , \\ 
  {\rm Multiple \ Stellar \ System:} \  B_{\rm amb} \, \lesssim \, 10^{-13}\, (n/1\cm )^{-2/3} \ {\rm G}. 
\end{array}
\right.
\]
Thus, the effect of the magnetic field can be ignored when the magnetic field in the ambient medium is weaker than 0.1\,pG, $B_{\rm amb} \ll 10^{-13}\,(n/1\cm)^{-2/3}$ G. 
Note that this condition may differ somewhat if the initial cloud  has a larger initial rotational energy.
\\

\item {\bf Mass Accretion Rate onto Primary Star}\\
In previous studies with one-zone or one-dimensional calculations, the mass accretion rate onto the Population III star was estimated as $\dot{M}\sim 10^{-2}-10^{-3} \msun$\,yr$^{-1}$.
In our calculation, all the models have almost the same mass accretion rate of $\dot{M}\sim 10^{-2} \msun$\,yr$^{-1}$.
Thus, the mass accretion rate derived in our three-dimensional calculation is comparable to that in spherically symmetric calculations.
In addition, the mass accretion rate depends little on the magnetic field strength.
When the primordial cloud has a relatively strong magnetic field,  no disk forms, and gas falls directly onto the protostar.
Thus, it is natural that the mass accretion rate in the magnetized cloud  corresponds well to that derived in spherically symmetric calculations. 
On the other hand, the gas accretion process is complicated in clouds without a magnetic field (or with an extremely weak field).
In such a cloud,  gas first falls onto the region near the protostar and tries to form a circumstellar disk.
However, fragmentation occurs just after the disk-like structure appears.
After disk fragmentation and clump formation, clumps disturb the circumstellar region.
The angular momentum around the primary star is effectively transferred by the gravitational torque caused by orbital motion of the clumps, and gas accretion onto the primary star is promoted.
%%In other words, the clumps acquire orbital angular momentum, whereas circumstellar gas lose its angular momentum and falls onto the central massive objects with the mass accretion rate of $\sim 10^{-2} \msun$\,yr$^{-1}$.
Although clump merger and mass exchange between the primary star and clumps contribute  to the mass growth of the primary star, the primary star effectively acquires most of its mass from the circumstellar region or the infalling envelope.
\\

\item {\bf No Stable Disk Formation and Final Stellar Mass}\\
In present-day star formation, a rotationally supported disk appears around the protostar.
In contrast, a persistent disk never appears in either magnetized or unmagnetized primordial clouds.
In unmagnetized primordial clouds, although a transient disk-like structure appears, fragmentation occurs, and several clumps form immediately after its formation.
The clumps absorb the gas in the disk and break up the disk.
In magnetized primordial clouds, no rotationally supported disk forms because the angular momentum is effectively transferred by magnetic effects (the magnetic braking catastrophe). 
Therefore, the formation process of Population III stars seems to differ from that of present-day stars.
However, we need a more long-term calculation with a more realistic setting to investigate further the evolution of Population III stars.
\end{itemize}

\section*{Acknowledgments}
We have benefited greatly from discussions with ~K. Omukai and ~H. Susa.
We are very grateful to an anonymous reviewer for a number of very useful suggestions and comments.
Numerical computations were carried out on NEC SX-9 at Center for Computational Astrophysics, CfCA, of National Astronomical Observatory of Japan.
%%This work was supported by Grants-in-Aid from MEXT (20540238, 21740136).

\clearpage
%%%%%%%%%%%%%
%%% Table1%%%
%%%%%%%%%%%%%
\begin{table}
\caption{Models and results}
\label{table:1}
%%\footnotesize
\begin{center}
%%\scalebox{.5}{%
\begin{tabular}{c|cccccc|cccc} \hline
{\footnotesize Model} & 
$B_0$ ($B_1$) [G]$^{*1}$ & $\Omega_0$  [s$^{-1}$] &  $\gamma_0 $ & $\beta_0$ & $\mu$ & Frag.$^{*2}$ & Jet$^{*3}$ \\
\hline
1  & 0 (0) & 0 & 0   & 0 & $\infty$ & No & No  \\
2  & 0  (0) & $8.5\times10^{-16}$ & 0   & $10^{-4}$ & $\infty$ & Yes & No  \\
3  & $10^{-5}$ ($2.2\times10^{-8}$) & $8.5\times10^{-16}$ & $2.6\times10^{-2}$   & $10^{-4}$ &6.4             &No & Yes  \\
4  & $10^{-6}$ ($2.2\times10^{-9}$) & $8.5\times10^{-16}$ & $2.6\times10^{-4}$   & $10^{-4}$ &64              &No & Yes  \\
5  & $10^{-7}$ ($2.2\times10^{-10}$) & $8.5\times10^{-16}$ & $2.6\times10^{-6}$   & $10^{-4}$ &640             &No & Yes  \\
6  & $10^{-8}$ ($2.2\times10^{-11}$) & $8.5\times10^{-16}$ & $2.6\times10^{-8}$   & $10^{-4}$ &$6.4\times10^3$ &No & No  \\
7  & $10^{-9}$ ($2.2\times10^{-12}$) & $8.5\times10^{-16}$ & $2.6\times10^{-10}$  & $10^{-4}$ &$6.4\times10^4$ & Yes (B) & No  \\
8  & $10^{-10}$($2.2\times10^{-13}$) & $8.5\times10^{-16}$ & $2.6\times10^{-12}$  & $10^{-4}$ &$6.4\times10^5$ & Yes & No  \\
9  & $10^{-8}$ ($2.2\times10^{-11}$) & $2.7\times10^{-15}$ & $2.6\times10^{-8}$   & $10^{-3}$ &$6.4\times10^3$ & Yes (B) & No  \\
10  & $10^{-9}$ ($2.2\times10^{-12}$) & $2.7\times10^{-15}$ & $2.6\times10^{-10}$  & $10^{-3}$ &$6.4\times10^4$ & No & No  \\
11 & $10^{-10}$ ($2.2\times10^{-13}$) & $2.7\times10^{-15}$ & $2.6\times10^{-12}$  & $10^{-3}$ &$6.4\times10^5$ & Yes & No  \\

12 & $10^{-5}$ ($2.2\times10^{-8}$) & $8.5\times10^{-15}$ & $2.6\times10^{-2}$  & $10^{-2}$ &$6.4$ & No & Yes  \\
13 & $10^{-6}$ ($2.2\times10^{-9}$) & $8.5\times10^{-15}$ & $2.6\times10^{-4}$  & $10^{-2}$ &$64$ & No & Yes  \\
14 & 0 (0) & $8.5\times10^{-15}$ & 0 (0)  & $10^{-2}$ &  $\infty$ & Yes & No  \\

\hline
\end{tabular}

\end{center}
$*$1 $B_0$ and $B_1$ are the magnetic field strengths at $n=10^4\cm$ and $n=1\cm$, respectively. \\
$*$2 Whether fragmentation occurred; $B$  indicates a binary system. \\
$*$3 Whether a jet appeared.
\end{table}

\clearpage
%%%%%%%%%%
% Fig. 1 %
%%%%%%%%%%

%%%%%%%%%%
% Fig. 2 %
%%%%%%%%%%

\clearpage
%%%%%%%%%%
% Fig. 3 %
%%%%%%%%%%
\begin{figure}
\includegraphics[width=150mm]{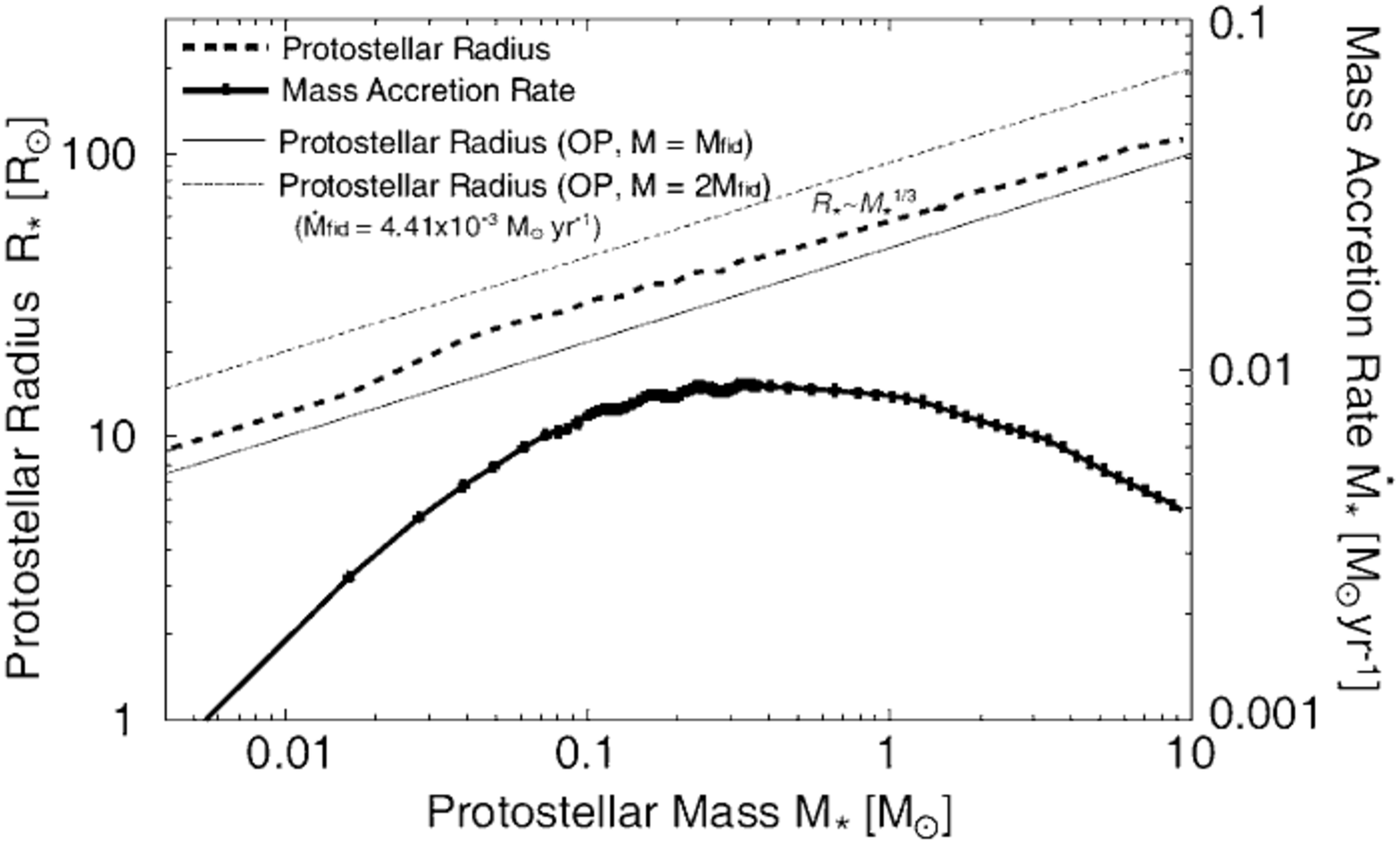}
\caption{
Protostellar radius (left axis) and mass accretion rate (right axis) against the protostellar mass.
Protostellar radii calculated  with $\dot{M}=4.41\times10^{-3}\msun$\,yr$^{-1}$ and $8.82\times10^{-3}\msun$\,yr$^{-1}$ by \citet{omukai03} are also plotted;  they are approximately determined by the relation $R_* \propto M_*^{1/3}$.
}
\label{fig:3}
\end{figure}

\clearpage
%%%%%%%%%%
% Fig. 4 %
%%%%%%%%%%

%%%%%%%%%%
% Fig. 5 %
%%%%%%%%%%

%%%%%%%%%%
% Fig. 6 %
%%%%%%%%%%

%%%%%%%%%%
% Fig. 7 %
%%%%%%%%%%

%%%%%%%%%%
% Fig. 8 %
%%%%%%%%%%

%%%%%%%%%%
% Fig. 9 %
%%%%%%%%%%

%%%%%%%%%%
% Fig.10 %
%%%%%%%%%%

%%%%%%%%%%
% Fig. 11 %
%%%%%%%%%%

%%%%%%%%%%
% Fig. 12 %
%%%%%%%%%%

%%%%%%%%%%
% Fig. 13 %
%%%%%%%%%%

%%%%%%%%%%
% Fig. 14 %
%%%%%%%%%%

%%%%%%%%%%
% Fig. 15 %
%%%%%%%%%%

%%%%%%%%%%
% Fig. 16 %
%%%%%%%%%%

%%%%%%%%%%
% Fig. 17 %
%%%%%%%%%%

%%%%%%%%%%
% Fig. 18 %
%%%%%%%%%%
\clearpage

\appendix
\section{Effect of Sink}
\label{sec:app1}
To investigate the effects of a sink cell on primordial star formation, we calculated the cloud evolution with a sink for models 1 and 2 of Table~\ref{table:1}.
We started the calculation without the sink and calculated the cloud evolution during the gas collapsing phase.
Later, we introduced the sink when the cloud density exceeded the threshold density $n_{\rm thr}$.
After sink creation, we removed the gas exceeding $n_{\rm thr}$ inside the accretion radius $r<r_{\rm acc}$ from the computational domain and added it to the protostar as a gravitating mass in each time step \citep[for a detailed description, see][]{machida10a,machida11a,machida12}.
To examine the effect of the sink, we prepared seven different criteria for sink creation by changing the threshold densities and accretion radii of the sink.
The sink models are listed in Table~\ref{table:a1}.
The Jeans length derived from the threshold density and corresponding temperature in our one-zone model is also listed.
In all models (S1-S6), the accretion radius is smaller than the Jeans length.
In the calculation, we resolved the accretion radius, using at least eight mesh points in the radial direction. 
%% which is determined  by the threshold density $n_{\rm thr}$ and corresponding gas temperature  \S\ref{sec:basic},
Thus, the Truelove condition \citep{truelove97} is satisfied during the calculation, and we can properly calculate  fragmentation with the sink.
%%The Jeans length is also listed in Table~\ref{table:1}.

\begin{figure}
\includegraphics[width=150mm]{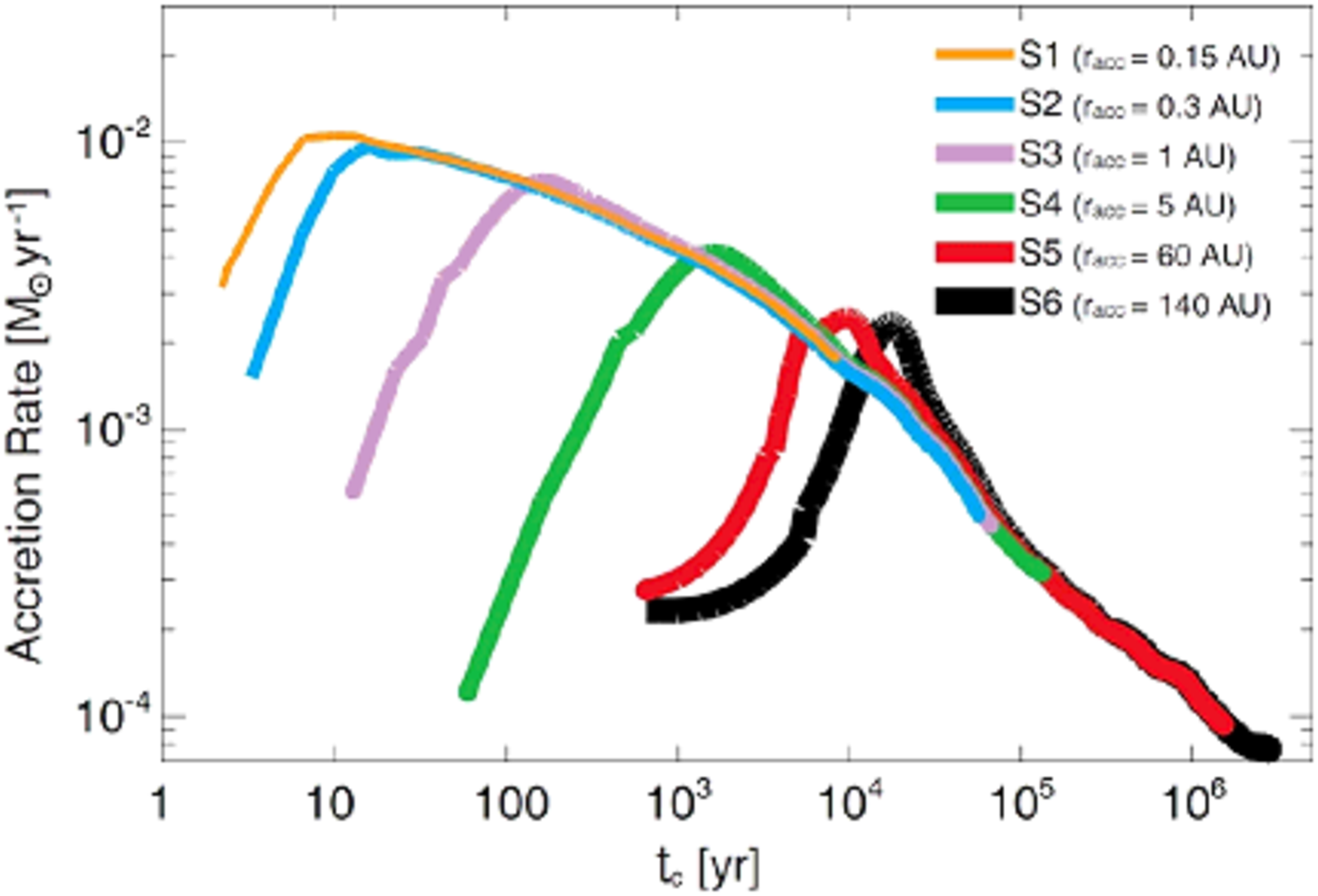}
\caption{
Mass accretion rate of non-rotating cloud (model 1) with different sink models (S1 -- S6) against 
the elapsed time after protostar formation.
}
\label{fig:a1}
\end{figure}

First, we calculated the evolution of a non-rotating cloud (model 1, see Table~\ref{table:1}) with the sink.
Figure~\ref{fig:a1} shows the mass accretion rate with different sink models against the elapsed time after sink creation (or protostar formation).
The figure indicates that the mass accretion rate depends very little on the accretion  radius (or threshold density) of the sink, although the cloud evolution just after protostar formation cannot be resolved when the accretion radius is very large (models S5 and S6).
In a non-rotating cloud, the gas simply falls onto the centre of the cloud, maintaining spherical symmetry without either the disk formation or jet emergence.
%% and thus the protostar does not affect its environment.
Therefore, the mass accretion rate depends only on the properties of the infalling envelope, which are determined during the gas collapsing phase before  protostar formation.
As a result, it seems that protostellar evolution in a non-rotating cloud can be calculated with a (large) sink.
In reality, radiation from the protostar should weaken the mass accretion  \citep{hosokawa11a,hosokawa12} when the protostar becomes sufficiently massive \citep{omukai10}.

\begin{figure}
\includegraphics[width=150mm]{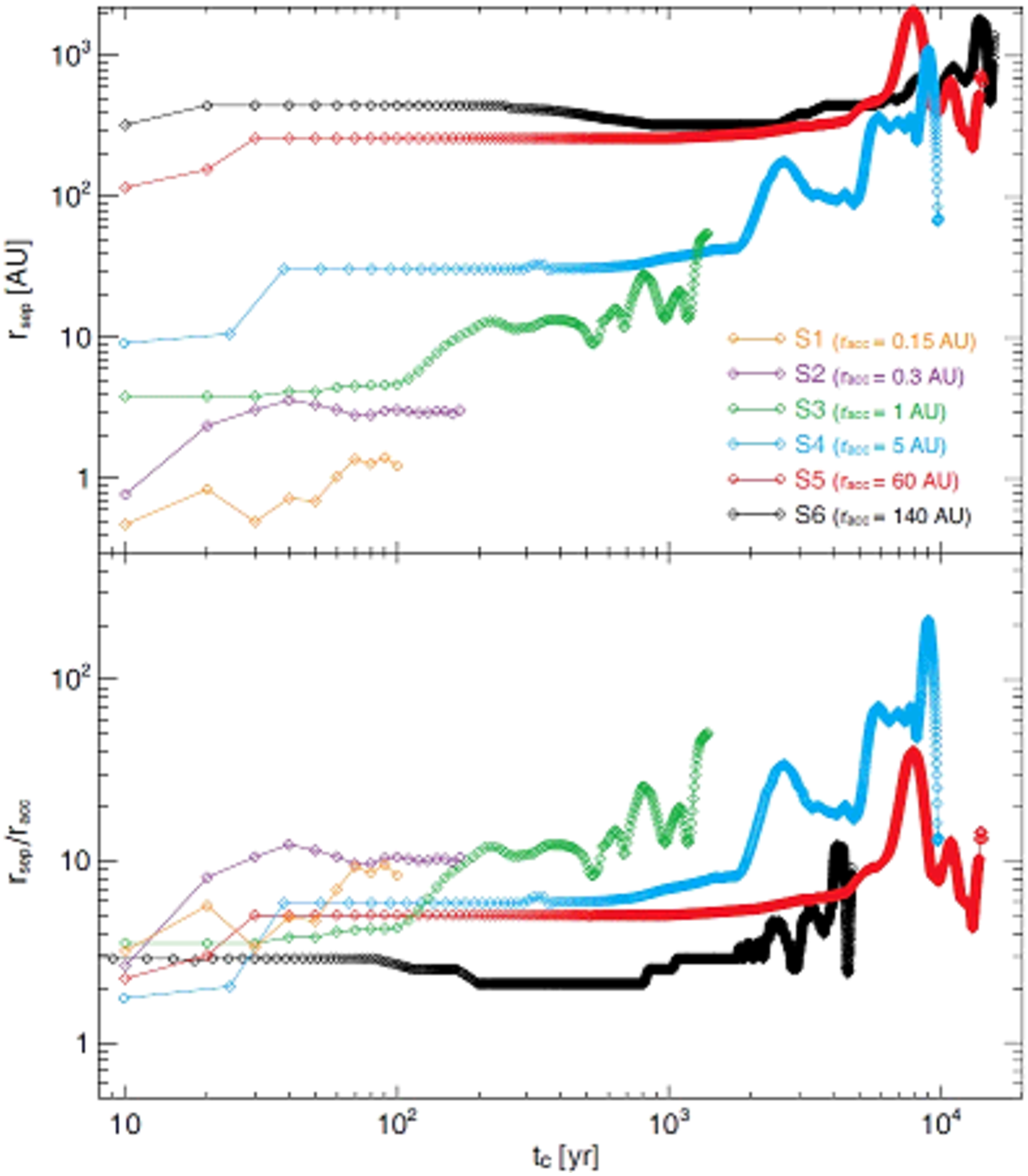}
\caption{
Top: Separations between primary and secondary stars against elapsed time after protostar formation for different sink models S1 -- S6. 
Bottom: Separations normalized by the accretion radius (or sink radius).
}
\label{fig:a2}
\end{figure}

In contrast to the case of the model without rotation, the sink greatly affects the evolution of a rotating cloud.
We also calculated the evolution of the rotating cloud (model 2) with different sink criteria (sink models S1-S6).
In the calculations, fragmentation occurs in all the sink models, although the fragmentation scales differ among the models.
The upper panel of Figure~\ref{fig:a2} shows the time evolution of the separation between the primary and secondary stars.
The separation of the fragments depends strongly on the accretion radius.
%%although we calculated the evolution of the same cloud, .
For example, the model with $r_{\rm acc}=140$\,AU (model S6) has a separation of $r_{\rm sep} \sim400$\,AU at $\tc=100$\,yr, whereas that with $r_{\rm acc}=0.15$\,AU (Model S1) has $r_{\rm sep} \sim 1$\,AU at the same epoch.

The lower panel of Figure~\ref{fig:a2} plots the separations normalized by the accretion radius and shows that fragmentation scale is 2-4 times the accretion radius.
Thus, Figure~\ref{fig:a2} indicates that fragmentation occurs at the scale of the sink (i.e. the accretion radius $r_{\rm acc}$) and the separation of the fragments in further evolutionary stages is also proportional to the accretion radius.
In summary, the evolution of a rotating cloud is controlled by the sink properties.

\begin{figure}
\includegraphics[width=150mm]{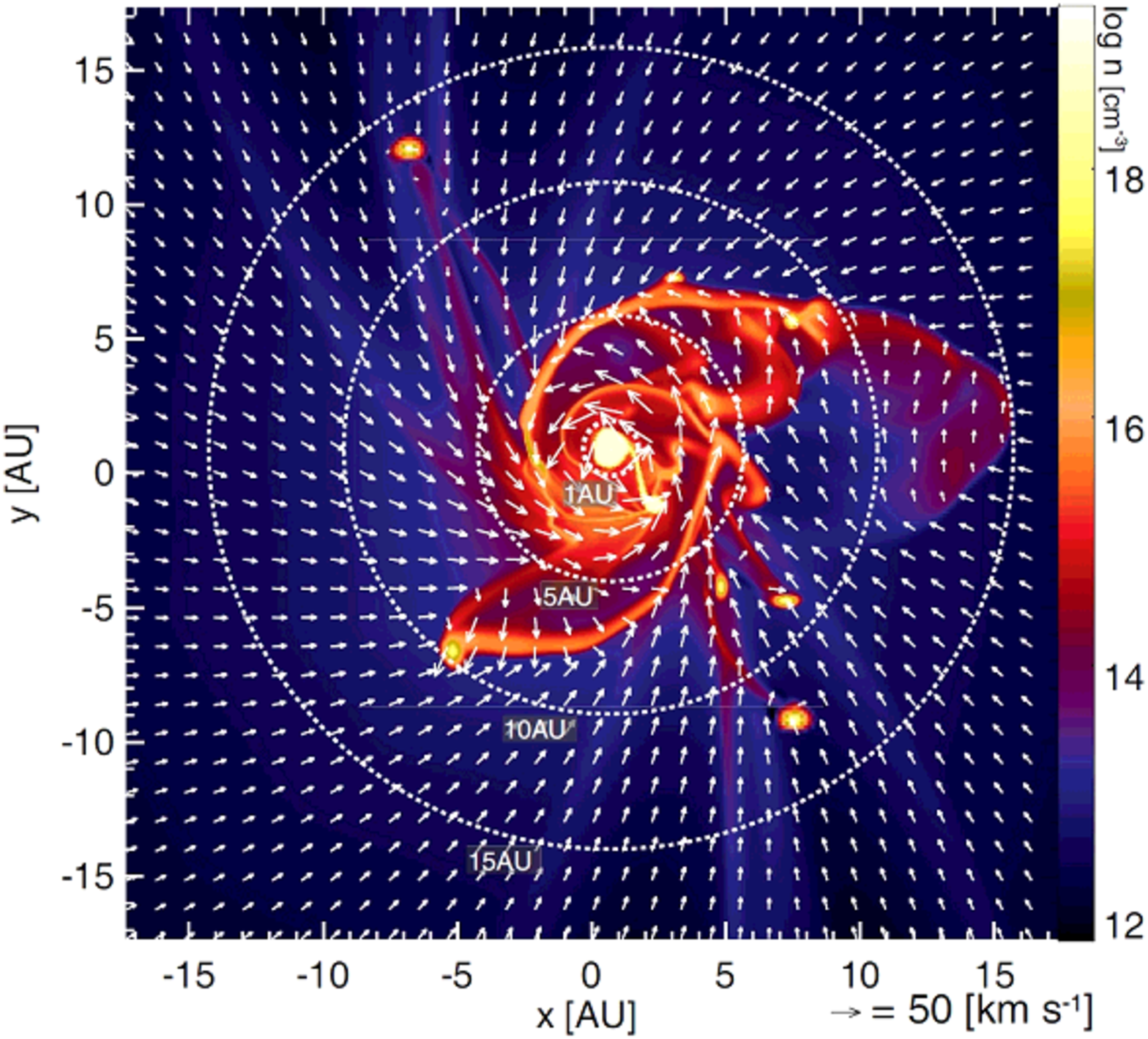}
\caption{
Density (colour) and velocity (arrows) distributions for model 2 at $\tc=301$\,yr.
Dotted circles indicate radii of 1, 5, 10 and 15\,AU from the primary (or first formed) protostar.
}
\label{fig:a3}
\end{figure}

Figure~\ref{fig:a3} shows the calculation result without the sink for model 2 and indicates that fragments are distributed at distances of $3-14$\,AU from the primary star. 
During the calculation, fragmentation occurs at $0.1\,{\rm AU} < r < 30\,{\rm AU}$.
In present-day star formation, fragmentation occurs at a scale of $\sim1-100$\,AU; this corresponds to the scale of the (rotating) first adiabatic core, which yield the typical fragmentation scale.
On the other hand, a primordial cloud has no typical scale, indicating that fragmentation is possible at any scale.
When the sink is adopted, the accretion radius becomes a typical fragmentation scale, and fragmentation occurs depending on the sink scale (or accretion radius). 
When no sink is used, the protostar gives the typical fragmentation scale, and fragmentation occurs near the protostar.
In principle, we should resolve the typical scale determined by the physics.
Therefore, we have to resolve the first adiabatic core, with a size of $\sim1$\,AU for present-day star formation and the protostar, with a size of $\sim 0.01$\,AU, for primordial star formation. 
A sink radius of $\sim1$\,AU may be applicable for present-day star formation, whereas a spatial resolution of $0.01$\,AU is necessary for primordial star formation.

%%%%%%%%%%%%%
%% Table A1%%
%%%%%%%%%%%%%
\begin{table}
\caption{Threshold density and accretion radius for sink}
\label{table:a1}
\begin{center}
\begin{tabular}{c|ccc} \hline
Model &  $ n_{\rm thr}$\ [$\cm$] & $r_{\rm acc}$\, [AU] & $\lambda_{\rm J}$\, [AU] \\ 
\hline
S1  & $10^{16}$ & 0.15 &0.95 \\
S2  & $10^{14}$ & 0.3  &10 \\
S3  & $10^{12}$ & 1    &86 \\
S4  & $10^{10}$ & 5    &900 \\
S5  & $10^{8}$ & 60    & 8200\\
S6  & $10^{7}$ & 140   & 22000\\
\hline
\end{tabular}
\end{center}
\end{table}

\clearpage
%%%%%%%%%%
% Fig. A1 %
%%%%%%%%%%

%%%%%%%%%%
% Fig. A2 %
%%%%%%%%%%

%%%%%%%%%%
% Fig. A3 %
%%%%%%%%%%

\end{document}